\documentclass[12pt]{article}
\usepackage{amssymb,amsmath,mathrsfs,amscd}
\DeclareMathOperator\arcsinh{arcsinh}
\begin{document}

%%%%%%%%%%%%%%%%%%%%%%%%%%%%%%%%%%%%%%%%%%%%%%%%%%%%%%%%%%%%%%%%

\begin{titlepage}

                            \begin{center}
                            \vspace*{1cm}
\large\bf{Entropies from coarse-graining: convex polytopes vs. ellipsoids}\\

                            \vspace{2.5cm}

              \normalsize\sf    NIKOS \  \ KALOGEROPOULOS $^\ast$\\

                            \vspace{0.2cm}
                            
 \normalsize\sf Weill Cornell Medical College in Qatar\\
 Education City,  P.O.  Box 24144\\
 Doha, Qatar\\

                            \end{center}

                            \vspace{2.5cm}

                     \centerline{\normalsize\bf Abstract}
                     
                           \vspace{3mm}
                     
\normalsize\rm\setlength{\baselineskip}{18pt} 

\noindent We examine the Boltzmann/Gibbs/Shannon $\mathcal{S}_{BGS}$  and 
the non-additive  Havrda-Charv\'{a}t / Dar\'{o}czy/Cressie-Read/Tsallis \ $\mathcal{S}_q$ \ and the Kaniadakis $\kappa$-entropy \ $\mathcal{S}_\kappa$ \  
from the viewpoint of coarse-graining, symplectic capacities and convexity. We argue that the functional form of such entropies can be ascribed to  
a discordance in phase-space coarse-graining between two generally different approaches: the Euclidean/Riemannian metric one that reflects independence 
and picks cubes as the fundamental cells and the symplectic/canonical one that picks spheres/ellipsoids for this role. 
Our discussion is motivated by and confined to the behaviour of Hamiltonian systems of many degrees of freedom.       
We see that  Dvoretzky's theorem provides asymptotic estimates for the minimal dimension beyond which these two approaches are close 
to each other. We state and speculate about the role that dualities may play in this viewpoint.\\   

                           \vfill

%\noindent\sf  PACS: \  \  \  \  \  \           \\
                                              \hspace*{20mm} Convexity, Symplectic capacities, Dvoretzky's theorem.   \\
                             
                             \vfill

\noindent\rule{8cm}{0.2mm}\\  
   \noindent $^\ast$ \small\rm E-mail: \ \  nik2011@qatar-med.cornell.edu\\

\end{titlepage}
 
%%%%%%%%%%%%%%%%%%%%%%%%%%%%%%%%%%%%%%%%%%%%%%%%%%%%%%%%%%%%%%%%%%

                                                                                \newpage                 
\setlength{\baselineskip}{18pt}

 \section{\large \  Introduction}

                                                                           % \vspace{5mm}

Entropy is one of the central concepts in Statistical Mechanics. Although initially introduced in thermodynamics, 
it has clearly superseded its modest origins and is currently widely used in numerous fields extending 
from dynamical systems and geometry to communication theory, complexity and far beyond. Due to its significance, considerable effort has 
been invested for almost 150 years, since its introduction by R. Clausius (ca. $\sim$1865) in understanding its meaning and ways to calculate 
it for specific systems. Fundamental contributions and interpretations, in various contexts, were made by  L. Boltzmann, J.W. Gibbs, J. von Neumann, 
C.Shannon, A. Kolmogorov, A. Renyi, I. Csiszar, E. Jaynes well as numerous other scientists and engineers ever since.  
The functional forms studied by most of the above people resemble each other very closely, so they tend to be treated together as one and the same entropy.
For the purposes of the present work alone we will pretend they are the same, although conceptually they are clearly very distinct from each other even in the 
specific context of Statistical Mechanics \cite{Bal-P, Bal-Book, Lesne}. 
Following this abusive bundling, we will be referring in the sequel to the Boltzmann / Gibbs / Shannon functional form  
\begin{equation}
       \mathcal{S}_{BGS} \ = \ - k_B \  \sum_{i\in I} \ p_i  \ \log p_i
\end{equation}
where $k_B$ is a constant which is identified in Statistical Physics as Boltzmann's constant. In (1) $i$ is an index taking values in a finite 
cardinality or countable set $I$, which is used to label the probabilities of  possible outcomes. \\

As is well-known, (1) has a  subjective character in Classical Physics \cite{Bal-P, Bal-Book}. The probabilities $p_i$ appearing in (1) depend not only on the 
actual system under study but also on the level of ignorance about the system by an experimenter/observer.  
In many occasions, it is worthwhile to consider systems with continuous rather than with discrete sets of outcomes. 
In such cases, we naively translate the definition (1) into the continuum, in which $p_i$ morph into a probability 
density function $\rho : \mathfrak{M} \rightarrow \mathbf{R}_+$. The details and the exact way of considering such a continuum limit may 
be a highly non-trivial process, during which one may have to introduce  a metric or a homogeneous structure \cite{Bal-P, Bal-Book} etc 
in order to reach a well-defined, and unique, result. Ignoring such subtle and important issues, the one would naively get
\begin{equation}    
         \mathcal{S}_{BGS} \ = \ - k_B \  \int_\mathfrak{M} \ \rho(x) \ \log \rho(x) \ d\mu
\end{equation} 
where $\mu$ is an appropriate measure on the sample space $\mathfrak{M}$. In the  case of Hamiltonian systems of many degrees of freedom, 
which is our object of study, $\mathfrak{M}$ usually stands for the phase space of the system, endowed with a Riemannian metric $\mathfrak{g}$ with 
$\mu$ being the unique Riemannian measure associated to $\mathfrak{g}$. \\

 There are several problems with (2) though: one is that it is not coordinate independent \cite{Bal-P, Bal-Book, Lesne}. 
This can be traced back to the fact that, very much like (1) from which it was naively inferred, (2) does not depend on the graph of $\rho$, but just on its range
of values. The functional (2) is not coordinate independent (diffeomorphism invariant).    
A second objection, related to the above comments, is that taking naively the limit of discrete $p_i$ to the continuum $\rho$ gives a divergent expression, the 
infinite part of which has to be judiciously subtracted before (2) can be properly used or interpreted.\\

Fortunately, using a relative entropy expression, \'{a} la Kullback-Leibler 
for instance,  addresses the first problem. So what one actually  computes in classical Statistical Mechanics is not really the ``absolute" entropy 
but rather a form of a relative entropy of a probability distribution with respect to an underlying background measure. 
If such a reference measure is independent of the details of the particular model at hand, then it results in
an additive constant which eventually becomes irrelevant as almost all experimentally verifiable quantities involve entropy variations rather than 
the ``absolute" value of the entropy itself.  
One chooses as a reference measure the probability density function resulting from the continuum 
limit of a discretization, usually  by cubes of side length $\sqrt{\hbar}$ of the phase space $\mathfrak{M}$.\\ 

One could question some of these statements, presenting in a counter-argument the case of the   Sackur-Tetrode equation which gives an 
expression for the (absolute) entropy of a classical monatomic ideal gas  
\begin{equation}
    \mathcal{S}_{BGS} \ = \ \frac{5}{2} k_B N + k_B N \log \left\{ \frac{m}{3\pi h^2}   \frac{V}{N} \left( \frac{U}{N}   \right)^\frac{3}{2}  \right\}  
\end{equation}
where $U$ is the internal energy of the gas, $m$ the mass of each atom, $V$ the volume that the gas occupies and $N$ the number of atoms. It seems that 
(3) contradicts the above statements pertaining to the additive constant. However the appearance in (3) of $h$, which is an arbitrarily chosen regularisation 
parameter in classical Physics and  only acquires physical significance as Planck's constant in Quantum Physics  
reinforces, rather than contradicts, the above statements. The implicit phase space discretisation by cubes of side $\sqrt{\hbar}$ appearing in expressions such as 
(3)  plays an important part in the present work. Its implicit, therefore often forgotten, presence is also at the heart of a recent controversy, to be mentioned below,
about the existence, appropriate form and physical significance of the continuum limit of the non-additive entropy $\mathcal{S}_q$.   \\

 As seen in the previous paragraph, in order to overcome the infinities that inevitably creep in the transition from (1) to (2) we can subtract a renormalisation 
 constant. Alternatively we can turn to Quantum Physics  and using the non Neumann operator functional 
\begin{equation}
     \mathcal{S}_{vN} \ = \ -  tr \ (\hat{\rho} \ \log \hat{\rho} )
\end{equation}
where $\hat{\rho}$ stands of the density operator (matrix) on the (kinematic) Hilbert space of the wave-functions and $tr$ is the trace over a basis of 
such a Hilbert space. The von Neumann entropy may still need some form of regularisation and renormalisation, especially in Quantum Field Theory. 
Its fundamental ``drawback" is that one has to know the underlying Quantum Theory, or at least to have a field theory description of  
the effective degrees of freedom before it can be properly implemented. 
An underlying quantum theory may not be known, as in the case of a quantum theory of gravity, 
for instance. We would  not want to even start a discussion about the technical obstacles of actually computing 
(3) for particular systems.  However, becoming aware of such well-known and long-ago resolved issues pertaining to $\mathcal{S}_{BGS}$ 
shows the subtlety and care with which the concept of entropy has to be treated.  \\

An additional incentive for looking into fundamental issues pertaining to entropy comes with the great success of $\mathcal{S}_{BGS}$ in quantifying the 
thermodynamic properties of systems at equilibrium. Such a success is seen in the agreement of the predictions derived by using $\mathcal{S}_{BGS}$ 
with numerous experimental observations, since the formulation of $\mathcal{S}_{BGS}$. But what is the source of such a spectacular success of 
$\mathcal{S}_{BGS}$? The dictum that we use $\mathcal{S}_{BGS}$  ``because it works" or it is the ``cannon"  
cannot be possibly satisfying, if someone is interested in getting a better understanding why things work the way they do. 
It is not obvious, for instance, why or whether  $\mathcal{S}_{BGS}$ precisely describes the collective behaviour  of systems with
long-range interactions, non-ergodic phase space evolution etc., not to even mention systems out of equilibrium.\\

 To understand the limitations of a particular functional one can compare its properties, predictions etc with those of another 
judiciously chosen or intelligently constructed functional. The Havrda-Charv\'{a}t/Dar\'{o}czy/Cressie-Read/Tsallis  entropy \ $\mathcal{S}_q$ 
\cite{HC, Dar, CR, RC, T1, T-book}      or the (Kaniadakis) 
$\kappa-$ entropy $\mathcal{S}_\kappa$ \cite{Kan1, Kan2, KanScar, Kan3, Kan4}, 
both to be defined and used below as ``alternative" (meaning in different regimes, or for different systems) functionals to $\mathcal{S}_{BGS}$.  
In addition, numerous other entropic functionals that have been recently introduced and used in Statistical Mechanics  \cite{T-book}, and more particularly the 
generalised exponential families specific members of which are $\mathcal{S}_q$ and  $\mathcal{S}_\kappa$ which were analysed in 
\cite{Nau1, Nau2, Nau-book},  can also be considered as  playing, in some part, such a role \cite{Nau-book}. 
A better understanding of such functionals and the determination of what are the essential physical features of the systems  whose collective behaviour they 
describe will not only help appreciate their significance but also set some boundaries to the complete dominance of $\mathcal{S}_{BGS}$ in Statisical Mechanics.
Therefore, such an effort it will also help us understand better $\mathcal{S}_{BGS}$ itself. \\

One of the many, still unanswered, questions pertaining to \ $\mathcal{S}_q$, \ $\mathcal{S}_\kappa$ and the numerous other non-additive entropic 
functionals is their dynamical basis. 
If \ $\mathcal{S}_{BGS}$ \ successfully describes systems having an ergodic evolution in their configuration or phase space, then what are 
the underlying dynamical features of systems, if any, described by \ $\mathcal{S}_q$, \ $\mathcal{S}_\kappa$ etc? 
The present work is partly motivated by and echoes to some extent, the general viewpoint described in as well as some of the fundamental issues 
pointed out in \cite{Coh}. Even though \cite{Coh} was written more than a decade ago, and despite the intervening considerable activity in 
understanding aspects of $\mathcal{S}_q$, \ $\mathcal{S}_\kappa$ and other non-additive entropic functionals, 
it is probably fair to state that most of the fundamental dynamical and statistical questions that \cite{Coh} had pinpointed remain unclear to this date. \\

In an attempt to address such questions, we explored  in our relatively recent work \cite{NK1, NK2, NK3, NK4, NK5, NK6, NK7, NK8, NK9}, 
some formal consequences of the definition of 
\ $\mathcal{S}_q$. \ At no point however  did we deal in any part of these works with the actual nature of $\mathcal{S}_q$ per se. 
We just confined ourselves to formal algebraic and geometric structures and conclusions stemming from its functional form.         
One of our key assumptions  was that some of the algebraic properties of \ $\mathcal{S}_q$ \ are not emergent from statistical averaging, 
but they directly reflect dynamical properties of the phase space of the system. 
In other words, we assumed that such algebraic properties of \ $\mathcal{S}_q$ \ are ``typical" of the underlying Hamiltonian system whose statistical 
behaviour is described by $\mathcal{S}_q$. A part of the present work is to investigate this assumed ``typical" 
behaviour and try to determine how it may dictate, at least some parts of the functional form of the entropy used to describe such systems. \\  

At the core of the present work is a question that appears trivial at first sight: if one knows the microscopic evolution of a Hamiltonian system
of many degrees of freedom, can this person  predict, or pick among various ``reasonable"  entropic functionals a unique one or, to be less ambitious, 
 a class of entropies that would successfully describe the macroscopic behaviour of the system? 
The obvious answer appears to be negative, as statistics seems in the eyes of many to be completely  independent/dissociated  from the underlying dynamics. 
It seems that we can successfully do the former, as in the case of $\mathcal{S}_{BGS}$  without knowing almost anything about the latter.  
This is certainly the viewpoint advocated, among others, by J.W. Gibbs, L.D. Landau and A.I. Khintchin who consider the underlying dynamics 
to be largely irrelevant, inasmuch as the ergodic hypothesis can be used to justify the choice of the micro-canonical ensemble. In this viewpoint  
 the success of Statistical Mechanics is ascribed to the large numbers of degrees of freedom of such systems \cite{CFLV, FSV}. 
The quantification comes by the Central Limit Theorem which ``justifies" the ``ubiquity" of the Gaussians in physical, and not only, processes. 
However, one may wish to note that the Central Limit Theorem does not hold if the random variables are not independent or weakly correlated. 
When there are non-trivial (``strong") correlations among such variables the  question naturally arises as to which statistics, hence which entropic 
functional, is appropriate for describing systems having such properties. This also emphasises the question about the meaning of ``independence" 
and how it needs to be modified, if at all, for the cases of these ``different kinds" of statistics. \\   

By contrast, we follow the view of L. Boltzmann (in part), P. Ehrenfest and A. Einstein \cite{Coh2}
according to which the underlying dynamics is at the core of the thermodynamic behaviour of a system. 
We believe that the recent emergence of numerous entropic functionals and the explorations into the realm of non-ergodic 
evolutions in phase space, make the underlying dynamical explorations highly desirable and potentially enlightening. 
We view the emergence of an entropic functional form for a Hamiltonian system with many degrees of freedom  
as a manifestation of a dissonance in phase space: usually one coarse-grains \cite{Gor, CFLV, QQ, QL} phase space by using cubes of side 
length $\sqrt{\hbar}$. Such cubes however do not behave well under canonical transformations. Based on the symplectic non-squeezing theorem  and 
the subsequent formulation of symplectic capacities as fundamental constructions in symplectic geometry, it is probably more prudent to coarse-grain 
$\mathfrak{M}$ in terms of ellipsoids. This happens because all symplectic capacities have the same value on ellipsoids. 
Generalizing cubes into convex polyhedra to take into account the composition of the newer, non-additive, entropic forms, we 
can see such entropy as arising from the  difference in coarse-graining  between ellipsoids and convex polyhedra. 
We choose the Banach-Mazur distance to quantify such a difference. Hence the problem reduces to determining the Banach-Mazur distance between
polyhedra and ellipsoids in $\mathfrak{M}$ of typical side/radius length $\sqrt{\hbar}$. Since ellipsoids  are minimal from the viewpoint of 
dynamics / symplectic capacities but the polytopes do not have any a priori lower bound on their size, we will consider the distance between such
polytopes and the largest spheres/ellipsoids that can be inscribed in them. A central result in the asymptotic limit of large  $n$              
is provided by Dvoretzky's theorem, the lower bound in the dimension of which gives rise to the functional form of $\mathcal{S}_{BGS}$ and  
provides the leading asymptotic form for non-additive entropies.\\      
       
In Section 2, we present some of the properties of \ $\mathcal{S}_q$ \ and \ $\mathcal{S}_\kappa$ \ that we need in this work. 
In Section 3, we briefly discuss the geometry of ``independence" and aspects of phase space coarse-graining.   
In Section 4, we present background material about Hamiltonian systems and symplectic geometry. 
In Section 5, we present basics of convex geometry/analysis needed to follow our exposition. 
In Section 6, we discuss Dvoretzky's theorem and dimension and the role played by dualities in this viewpoint.
Section 7 contains conclusions and some speculations.  \\

%%%%%%%%%%%%%%%%%%%%%%%%%%%%%%%%%%%%%%%%%%%%%%%%%%%%%%%%%%%%%%%%%%%%%%%%%%

                                                                  \section{\large \ Structures induced by two non-additive entropies.}

                                                                             \vspace{3mm}
 
 Two of the most commonly used non-additive entropies, which have attracted substantial attention recently are presented in this Section. 
 Pertinent properties, for this work, are also stated. \\
                                                                  
%%%%%%%%%%%%%%%%%%%%%%                                                                  

\subsection{\normalsize The Tsallis entropy $\mathcal{S}_q$  and its induced operations.}

The Havrda-Charv\'{a}t \cite{HC}, Dar\'{o}czy \cite{Dar}, Cressie-Read \cite{CR, RC}, Tsallis \cite{T1, T-book} 
entropy  \ $\mathcal{S}_q$ \ introduced and developed, in part, in the context of Statistical Mechanics by Tsallis  
for a discrete set of outcomes $\{ p_i \}$, parametrized by the index set $I$ and with $i\in I$, is given by 
\begin{equation}
         \mathcal{S}_q [\{ p_i \} ] \ = \ k_B \ \frac{1}{q-1} \left\{ 1 - \sum_{i\in I} \  p_i^q \right\}  
\end{equation}
where $k_B$ is the Boltzmann constant. 
Its continuum analogue  for a sample space \ $\Omega$ \ equipped with a measure absolutely continuous with respect to the Lebesgue measure  
with Radon-Nikodym density \ $\rho$ \  is naively assumed to be  \cite{T1, T-book}
\begin{equation}      
   S_q [\rho]  \ =  \ k_B \ \frac{1}{q-1} \left\{1 - \int_{\Omega} [\rho(x)]^q \ dvol_{\Omega} \right\}
\end{equation}
Here \ $dvol_{\Omega}$ \ represents the infinitesimal volume element of \ $\Omega$ \ when it is a Riemannian manifold \ $\mathfrak{M}$, \ as is 
usually the case for the Hamitonian systems of many degrees of freedom which are the focus of our attention.  
It should be noted that most recently there has been a controversy  regarding the validity of this naive extension of 
$\mathcal{S}_q$ to continuous variables, without either side being definitively convincing, in our opinion \cite{Abe1, Andre, Abe2, BOT, BL, LB, QL, PR}.
The controversy brought about by \cite{Abe1} is intimately related to the implicit normalization of any entropy functional, such as 
$\mathcal{S}_{BGS}$ for instance, required to make its definition coordinate independent (diffeomorphism invariant). It is usually provided by the 
discretizaton of $\mathfrak{M}$ in cubes of side length $\sqrt{\hbar}$, of the density distribution arising from its continuum limit. For  $\mathcal{S}_{BGS}$   
due ore the presence of $\log$ it results in a constant that is additive and hence can be ignored inasmuch as entropy differences are the only relevant
quantities in physical predictions. By contrast, such a term is not additive, but rather multiplicative,  therefore cannot be omitted in (6) in considering 
entropy differences. Subsequently the above authors have presented their views on this and related matters that may make the use of (6) rather 
questionable. This matter is of interest but not of central importance in the line of arguments and viewpoint of the the present work, therefore  will 
sidestep these issues in what follows, and keep using (6) pretending that this naive is indeed a valid generalisation to the continuous case.\\   

The nonextensive parameter \ $q$ \ can generically take any values in \ $\mathbf{R}$. \  There has been a recent proposal to extend its validity to 
\ $q\in\mathbf{C}$ \ which is certainly worth looking into as well as considering the associated interpretation of such an extension \cite{WW}. 
To have desirable properties, such as relative insensitivity to rare events, convexity, and decay in a polynomial manner \cite{T-book}, and following our 
past work \cite{NK1, NK2,  NK3, NK4, NK5, NK6, NK7, NK8, NK9}  as well as the more recent \cite{BL, LB}, we will assume that \ 
$q\in [0,1] \subset \mathbf{R}$ everywhere  in the sequel. We straightforwardly notice, that for \ $q\rightarrow 1$ \ one recovers  $\mathcal{S}_{BGS}$. 
We will set henceforth \ $k_B = 1$ \ for brevity. \\

Conventionally, two subsystems \ $A, B \subset \Omega$ \ are considered independent \cite{Touch} if their marginal probability distribution functions 
are related by 
\begin{equation} 
     p_{A \cup B} = p_A p_B
\end{equation}
Here $A \cup B$ indicates the system resulting from the interaction of $A$ and $B$. 
For such subsystems $\mathcal{S}_{BGS}$ is easily seen to be additive, namely 
\begin{equation}
      \mathcal{S}_{BGS} (A\cup B)  \   =  \ \mathcal{S}_{BGS} (A) + \mathcal{S}_{BGS} (B)
\end{equation}
The  entropy $\mathcal{S}_q$ however  is not additive \cite{T-book}, at least not in the conventional sense, as it satisfies
 \begin{equation}   
  \mathcal{S}_q (A\cup B)  \  =  \  \mathcal{S}_q (A) + \mathcal{S}_q (B) + (1-q) \mathcal{S}_q(A) \mathcal{S}_q(B) 
 \end{equation}           
This lack of additivity is usually ascribed to the long-range spatial and temporal correlations of the systems that $\mathcal{S}_q$ entropy conjecturally 
describes \cite{T-book}. For systems described by such $\mathcal{S}_q$, additivity is manifestly restored if the addition is redefined as 
\cite{NLeMW, Borg}
\begin{equation}  
        x \oplus_q y \ = \ x + y + (1-q)xy     
\end{equation}            
Then 
\begin{equation}
      \mathcal{S}_q (A\cup B) \ = \ \mathcal{S}_q(A) \oplus_q \mathcal{S}_q(B)
\end{equation}    
It took sometime before a generalized product, distributive with respect to the addition (10) was discovered \cite{PLCPB, NK1} 
Even though \cite{PLCPB} and \cite{NK1} gave 
different forms of such a product, conjecturally equivalent, we will be using here the one introduced in \cite{PLCPB} as more elegant and easier  to work with. 
The generalised product turned out to be   
\begin{equation}
   x \otimes_q y = \frac{1}{1-q} \cdot  \left\{(2-q)^\frac{ \log[1+(1-q)x]  \log[1+(1-q)y]}{[\log (2-q)]^2} - 1\right\} 
\end{equation}
The definition of the generalised product (12) appears to be somewhat arcane. However, the motivation behind its construction becomes more 
transparent, in our opinion,  if we see it as a result of demanding the commutativity of the diagram \cite{NK2}
\begin{equation}
       \begin{CD}
         \mathbb{R} \times \mathbb{R}                            @ > \cdot >>                                  \mathbb{R} \\
                    @V \tau_q \times \tau_q VV                                                                              @VV \tau_q V\\
         \mathbb{R}_q\times\mathbb{R}_q                   @ >> \otimes_q \  >                          \mathbb{R}_q        
        \end{CD} 
   \end{equation}
Putting together the two generalized binary operations (10) and (12), we set up \cite{NK2, NK3} a one-parameter family of deformations of the set of reals 
denoted by  \ $\mathbb{R}_q$. \ An explicit isomorphism between these two sets  \ $\tau_q: \ \mathbb{R} \rightarrow \mathbb{R}_q$ \ is given by 
\cite{NK2, NK3}    
\begin{equation}
       \tau_q (x) \ =  \  \frac{(2-q)^x - 1}{1-q}, \ \ \  \ q\in [0,1) 
\end{equation}
In terms of this field isomorphism, the product (12) can be rewritten as 
\begin{equation}
     \tau_q (x\cdot y) \  =  \ x_q \otimes_q \ y_q  \ =   \  \tau_q \left( \tau^{-1}_q (x_q)\cdot  \tau^{-1}_q (y_q) \right)
\end{equation}
By the above construction, we have in effect reduced the differences between \ $\mathcal{S}_{BGS}$ \ and \ $\mathcal{S}_q$ \ to the differences 
between \ $\mathbb{R}$ \ and \ $\mathbb{R}_q$. \ This can more formally seen through a comparison between the axioms used to determine \ 
$\mathcal{S}_{BGS}$ \ \cite{Shan, Khin} \ and  \  $\mathcal{S}_q$ \ \cite{Santos, Abe3}. \\    

In closing this Section, we should point out that there are several distinct and non-equivalent definitions of $q$-exponentials in the literature 
(see e.g. \cite{Kac}) quite frequently associated to quantum groups, which have nothing obvious to do with $\mathcal{S}_q$.  
With eyes to the next subsection, the same words of caution also apply to the several $\kappa$- 
distributions existing in the literature (e.g. \cite{Livad,PierLaz} in space plasmas) which have nothing obvious to do with $\mathcal{S}_\kappa$.  
Due to this lack of uniformly in nomenclature,  one should be careful about the exact functional forms are used in each occasion.  \\

%%%%%%%%%%%%%%%%%%%%%%%%%

\subsection{ \normalsize The $\kappa$-entropy \  $\mathcal{S}_\kappa$ \  and its induced operations.}

Among the many entropic functionals that have been constructed over the years, the $\kappa$-entropy 
$\mathcal{S}_\kappa$ has also attracted some attention since its introduction \cite{Kan1, Kan2, KanScar, Kan3, Kan4}. 
Unlike $\mathcal{S}_q$ whose origin can be traced 
to the thermodynamic formalism, the origins and possible scope of $\mathcal{S}_\kappa$ are far more concrete: they 
rely on attempts to understand the thermodynamic behaviour of the free relativistic gas, a system whose thermodynamic behaviour 
has proved to be far harder to describe than could be naively suspected. Since we do live in a relativistic world, where locally the 
principle of Relativity describes many physical phenomena, determining how it dictates the collective behaviour of systems of many 
degrees of freedom may be of considerable importance. \\

The $\kappa$-entropy was introduced as a functional that generates through the variational principle, a given the $\kappa$-exponential
distribution that arose from arguments pertaining to non-linear kinetics \cite{Kan1}.  Lorentz invariance is already built into the underlying dynamics in 
this formalism. The $\kappa$-entropy was defined directly for a continuous probability distribution with density $\rho$ on the sample space 
$\Omega$ as \cite{Kan1, KanScar} 
\begin{equation} 
    \mathcal{S}_\kappa [\rho ] \ = \  \int_\Omega  \left\{ c(\kappa ) \rho^{1+\kappa } + c(-\kappa ) \rho^{1-\kappa} \right\} \ dvol_\Omega,   
                                                               \hspace{10mm}  c(\kappa) \ = \  - \frac{Z^\kappa}{2\kappa (1+\kappa )}   
\end{equation}
where $Z$ is a normalisation constant and $\kappa\in\mathbf{R}$. So far as the author knows, there is no standing proposal to extend 
$\kappa\in\mathbf{C}$ although we do not see any reason what this would not be feasible, if the need arose and an appropriate 
physical motivation and interpretation would be provided as in the case of $\mathcal{S}_q$. We expect about this continuous 
functional form worries/objections similar to the ones that arose for $\mathcal{S}_q$ which were alluded to above. The discrete analogue of    
(16) for a set of outcomes $I$ with corresponding probabilities $\{ p_i \}, \ i\in I$ would naively appear to be 
\begin{equation}
    \mathcal{S}_\kappa [\{ p_i \} ] \ =  \ \sum_{i\in I}  \left\{ c(\kappa ) p_i^{1+ \kappa } + c(-\kappa ) p_i^{1-\kappa } \right\}  
\end{equation}
with $c(\kappa )$ as in (16). One should be quite careful though in providing such naive discrete generalization, if one is interested in 
maintaining for (17) some form of Lorentz-invariance as the one that a gave rise to (16). It is well-known, and probably obvious,  that discrete 
structures violate manifest Lorentz-invariance, something that has presented major technical challenges to proponents of quantum gravity 
theories. The solutions, which could also be adopted here, is to either forego completely any requirements for even remnants of Lorentz-invariance in (17), 
or to use arguments relying on randomness that preserve such a structure, as was done, for instance, in \cite{BomHenSor} for the case of causal sets.\\      

One can immediately observe that 
\begin{equation}
     \lim_{\kappa \rightarrow 0} \mathcal{S}_\kappa \ = \ \mathcal{S}_{BGS} 
\end{equation} 
It is also immediately obvious that $\mathcal{S}_\kappa$ are not additive with respect to the usual addition and that to restore manifest additivity
one will leave to define the generalised sum as \cite{KanScar, Kan4} 
\begin{equation}
      x \stackrel{\small \kappa}{\oplus} y \ = \ x \sqrt{1+\kappa^2 y^2} \ + \ y \sqrt{1+\kappa^2 x^2} 
\end{equation}
where $|\kappa | < 1$, mirroring  (10). This can be re-written as 
\begin{equation}
      x \stackrel{\small\kappa}{\oplus} y \ = \  \frac{1}{\kappa} \sinh \left( \arcsinh (\kappa x) + \arcsinh (\kappa y) \right)
\end{equation}
to resemble more closely the generalized product, to be defined next. 
The generalised product \cite{KanScar, Kan4}, which is distributive with respect to the generalised sum (19) and mirroring (12)  
turns out to be \cite{KanScar, Kan4}
\begin{equation}
     x \stackrel{\small \kappa}{\otimes} y \ = \ \frac{1}{\kappa} \sinh \left( \frac{1}{\kappa}  \ \arcsinh (\kappa x) \  \arcsinh (\kappa y ) \right)
\end{equation}
Then, in parallel to the case of $\mathcal{S}_q$,  one \cite{KanScar, Kan4} can define the deformed field 
$\mathbf{R}_\kappa = (\mathbf{R}, \stackrel{\small\kappa}{\oplus}, \stackrel{\small\kappa}{\otimes})$ 
and set set up \cite{KanScar, Kan4} a field isomorphism $\tau_\kappa: \mathbf{R} \rightarrow \mathbf{R}_\kappa$ which is given explicitly by 
\begin{equation} 
      \tau_\kappa (x) \ = \ \frac{1}{\kappa} \arcsinh (\kappa x)  
\end{equation}
with an inverse
\begin{equation}
      \tau_\kappa^{-1} (x) \ = \ \frac{1}{\kappa} \sinh (\kappa x)
\end{equation}
in analogy with (14) for $\mathcal{S}_q$ and also mimicking (15) we get
\begin{equation}
      \tau_\kappa (x\cdot y) \ = \ x_{\kappa} \stackrel{\small\kappa}{\otimes} y_\kappa \ = \ 
                                             \tau_\kappa \left( \tau_\kappa^{-1} (x_\kappa) \cdot \tau_\kappa^{-1} (y_\kappa) \right)
\end{equation}
We are not aware of an existing axiomatic formulation of the $\kappa$-entropy, 
but we do not consider this as a drawback at a physical level, but rather as an open question at the formal level that 
remains to be addressed, if such interest arises, in the future.  \\

%%%%%%%%%%%%%%%%%%%%%%%%%%%%%%

\subsection{\normalsize Features of $\mathcal{S}_q$ and $\mathcal{S}_\kappa$ entropies.}

We observe from the above structures, that even though $\mathcal{S}_q$ and $\mathcal{S}_\kappa$ are, arguably, the two most developed 
non-additive entropic functionals in Statistical Mechanics to date, they are not really all that different from each other in terms of their induced 
structures. For instance, one can easily argue as in \cite{NK3, NK10}, that $\mathcal{S}_\kappa$ also describes cases of ``weak chaos". 
To be more precise \cite{NK3, NK10}, one can straightforwardly see that if a system is described by the $\kappa$-entropy 
and the composition (20) or (21) is a reflection of its phase space metric properties, 
rather then being an emergent property due to the statistical averaging, then the 
largest Lyapunov exponent of the underlying dynamical system will be zero, in complete analogy with the case of $\mathcal{S}_q$.   
This commonality can be traced back to the similarity between the functional forms of $\mathcal{S}_q$ and $\mathcal{S}_\kappa$: 
both (6) and (16) can be seen to have a functional form that is asymptotically exponential. These functional forms are actually 
suggestive of the different parametrizations of the hyperbolic space  \cite{CFKP}. 
Of course, this does not mean that the actual functional forms of $\mathcal{S}_q$ and $\mathcal{S}_\kappa$ are the same, or that they 
will give rise to the same physical predictions, but they should share asymptotical, some  common features such as describing weak chaos. 
It would be of great interest to compare the features of the systems that  are described be each one of these two entropic
functionals. We believe that someone should be able to say some similar things for many, if not necessarily all, probability distributions 
belonging to the exponential family, aspects of which have been developed in \cite{Nau1, Nau2, Nau-book}. \\    

The non-uniqueness of $\mathcal{S}_q$, at least from the viewpoint of its composition properties, but the fact that it is a part of a larger family 
of functional forms that share many common features was also briefly touched upon in \cite{NK9}. It was noticed in \cite{NK9} that even though
$\mathcal{S}_q$ was an interesting case of a functional form belonging to the displacement convexity class $\mathcal{DC}_N$ for 
\begin{equation} 
     N \ = \  \frac{1}{1-q}
\end{equation}
it was not unique, by any means.  Its uniqueness was restored, when in addition to (10), one could invoke the other axioms of \cite{Santos, Abe3}. 
It is not obvious, to us  at least,  that some of these axioms, even if reasonable, should necessarily describe the properties of the entropy  for 
systems out of equilibrium, with long range temporal and spatial correlations, etc. 
In accordance with the functional forms of the generalized entropies used in defining the Bakry-\'{E}mery-Ricci curvature thorough optimal 
transportation, as  presented in \cite{NK9}, it may be more prudent to consider $\mathcal{S}_q$ as just one interesting example of an entropy 
having a polynomial/power-law form rather than as the unique entropy that may describe properties of the systems having properties that are 
mentioned above. Therefore the afore-mentioned interest in the analysis of systems that are described by one of such entropies, but not for the 
others, may help clarify their range of applicability or  even the physical mechanisms leading to their effectiveness in describing the macroscopic 
properties of such systems.\\     

%%%%%%%%%%%%%%%%%%%%%%%%%%%%%%%%%%%%%%%%%%%%%%%%%%%%%%%%%%%%%%%%%%%%%

                                                   \section{\large \ The ``shape" of independence and phase-space coarse-graining.}

                                                                               \vspace{3mm}

In this Section we analyse the concept of ``independence" and the subsequent shape it induces on the fundamental cells in phase-space coarse-graining,  
with a view toward $\mathcal{S}_{BGS}$ and the non-additive entropies of the previous Section. \\

%%%%%%%%%%%%%%%%%%%%%%%%

\subsection{\normalsize  Independence and cubes.}

 The conventionally accepted formalisation of the concept of independent interacting 
subsystems was stated in (7) and is realised through the multiplicative character of the marginal probability distributions
of the interacting subsystems. In the closely related case of random variables \ $X$ \ and \ $Y$, \  
they are called ``independent" if 
\begin{equation}
     \mathbb{E} [X\cdot Y] \ = \ \mathbb{E}[X] \cdot \mathbb{E}[Y]
\end{equation}
where \ $\mathbb{E}$ \ stands for ``expectation value" of the corresponding random variable \cite{Tal}. 
At the set-theoretical level, ``independence"  is conventionally encoded via the Cartesian product of sets. From this viewpoint, 
the simplest set expressing set-theoretic independence is the unit cube in \ $\mathbb{R}^n$ \ indicated by 
 \begin{equation}
 I_n \ =  \ [-1,1]^n 
 \end{equation}
From (7) it becomes obvious that the concept of probabilistic independence is intimately related to multiplicative-like structures \cite{Tal}.
Hence modifications in the definition of multiplication, as in (12), (21) for instance, will have significant implications for determining what constitutes 
``independent" outcomes. Through all this, we want to indicate that the introduction of (12) which was induced by \ $\mathcal{S}_q$ \ and (21) for 
\  $\mathcal{S}_\kappa$, \   
forces us to re-think and modify the concept of ``independence" in the framework of the non-additive entropies (6), (16). 
This modification of ``independence"  is necessary due to the long-range temporal and spatial correlations of the systems  that
the non-additive entropies describe. When such correlations are present, the conventional definition (7) does not behave well (``covariantly")
with respect to the structures induced by the underlying entropies such as $\mathcal{S}_q$, or  $\mathcal{S}_\kappa$. As stated above, if we want 
to assume that the macroscopic algebraic and geometric properties are a direct reflection of the microscopic dynamics, and not emergent due to 
statistics, then a more ``covariant'' definition of independence at the microscopic level would be     
\begin{equation}     
     p_{A\cup B} \ = \ p_A \otimes_q p_B 
\end{equation}
or 
\begin{equation} 
    p_{A \cup B} \ = \ p_A \stackrel{\small \kappa}{\otimes} p_B
\end{equation} 
following (12) or (21) respectively. Since there is no obvious generalisation of the Cartesian product in such cases, it is hard to see how one 
can find the counterparts of the unit cube (27)  for the generalised products (12), (21). \\

%%%%%%%%%%%%%%%%%%%%%%%%%%%%%%%%%%%%%%%%%%%%%%

\subsection{\normalsize Generalized independence and polytopes.}

The question that therefore naturally arises is how to determine such generalised ``cubes" whose shape would express generalized 
independence the same way that (27) expresses conventional independence. An answer is provided if one thinks of the cube in a metric, rather
than in a set-theoretic, way. Consider $\mathbf{R}^n$. Its elements $a$ are ordered $n$-tuples of real numbers $ a = (a_1, \ldots, a_n)$. 
Their $\mathbf{R}\ni p$-norm, for $p\geq 1$ so as the triangle inequality to be satisfied,  is defined by
\begin{equation}
   \| a \|_p \ = \ \left( \sum_{i=1}^n |a_i|^p \right)^\frac{1}{p} 
\end{equation}
where $| \cdot |$ stands of the absolute value of its argument. The sup-norm in $\mathbf{R}^n$ can be seen either as 
\begin{equation}
      \| a \|_\infty \ = \ \sup_{i=1,\ldots, n} \{ |a_i | \}
\end{equation}
or, equivalently, as the limit
\begin{equation}
     \| a \|_\infty \ = \ \lim_{p\rightarrow\infty} \| a \|_p
\end{equation}
With such norms, $\mathbf{R}^n$ is a Banach space, indicated as $l^n_p$ or $l^n_\infty$ respectively.  
The ball indicated by  $B_r (x)$ of radius $r$  centered at a point $x$ of a metric space $\mathfrak{X}$ 
with distance function $d$ is defined by
\begin{equation}  
    B_r(x) \ = \ \{y\in \mathfrak{X}: \ d(x,y) \leq r \}
\end{equation}
and the sphere $S_r(x)$ of radius $r$ is defined by 
\begin{equation}
  S_r(x) \ = \ \{ y\in \mathfrak{X}: \ d(x,y) = r \} 
\end{equation}
One can easily see that the cube is the unit ball of  $l^n_\infty$, namely 
\begin{equation}
      I_n \ = \ B^\infty_1(0)
\end{equation}
where the superscript explicitly denotes the sup-norm. The advantage of this viewpoint is that it can be carried over 
directly  to infinite dimensions, namely to the space of sequences $(a_1, \ldots, a_n, \ldots )$ with elements in 
$\mathbf{R}$, namely to the Banach space $l_p, \ p\in [1, \infty]$. Actually there are several such reasonable infinite dimensional 
limits, depending on one's goals, but we will not enter the details of this. Such infinite dimensional limits are useful if one wishes to 
be able to consider the ``thermodynamic limit" $n\rightarrow\infty$ at some stage of these calculations. Moreover, such definitions can be 
generalised to uncountable spaces, such as the Lebesgue spaces $L^p (\mathbf{R}^n)$ of $p$-integrable functions, to Orlicz,
Sobolev and even more general function spaces \cite{Triebel} that may be useful. \\

One can then use the generalized operations of the deformed fields $\mathbf{R}_q$ and $\mathbf{R}_\kappa$  instead those of 
the usual addition and multiplication to define the generalised cubes $\mathfrak{I}^q_n$ and $\mathfrak{I}^\kappa_n$ in exactly 
the same way as it was done for $\mathbf{R}^n$ in the previous paragraph. This is possible because of the presence of the field
isomoprhisms (14), (23) which being distance non-decreasing maps, they also preserve the order structure of $\mathbf{R}$.
Hence the induced topologies by the generalized operations of $\mathbf{R}_q$, $\mathbf{R}_\kappa$ are homeomorphic, 
the ordering of the elements of these sets is maintained, therefore the supremum has an unambiguous meaning etc.    
Given such definitions, the  polytopes playing the role of the cubes $\mathfrak{I}^q_n$, \ $\mathfrak{I}^\kappa_n$ for the 
generalized products (12), (21) respectively, can all be seen to be given by 
\begin{equation}  
   \mathfrak{I}^q_n \ = \ \tilde{\tau}_q (I_n)
\end{equation}
and 
\begin{equation}
   \mathfrak{I}^\kappa_n \ = \ \tilde{\tau}_\kappa (I_n)
\end{equation}
where the tilde $\sim $ denotes the $n$-dimensional extension of its underlying isomorphism. \\ 

%%%%%%%%%%%%%%%%%%%%%%%%%%%%%%%%%%%%

\subsection{\normalsize  Euclidean and dynamical aspects of coarse-graining with cubes.}

The definitions of such cubes are particularly important in the context of coarse-graining of the phase space 
\cite{CFLV, Gor, GorKOT, GorKarl, GellMH1, GellMH2}. Coarse-graining was introduced by P. Ehrenfest and T. Ehrenfest 
in an attempt to explain the origin of macroscopic irreversibility, in the face of   microscopic  reversible dynamics. 
Many of these ideas can be traced back to L. Boltzmann. One way to implement coarse-graining 
 is to divide the phase space of the microscopic dynamical system (Hamiltonian of may degrees of freedom, in the case of our interest)
in cells and substitute the smooth probability density $\rho$ of phase space by a piece-wise 
constant one $\rho_{cg}$  in each of these cells. The size of each cell is assumed to be small  but it should not approach zero. 
For Boltzmann's ideas about the behaviour of  gases and the Sackur-Tetrode equation (3) the side length 
of each cube is taken to be $\sqrt{\hbar}$. Effectively what this approach to coarse-graining does is to combine elements of the 
microscopic evolution of the system with a periodic partial equilibration. The end result is to determine a macroscopic kinetic 
equation that does not retain any memory of the initial condition of the system but captures the evolution of these successive 
partial equilibrations \cite{Gor, GorKOT, GorKarl}.\\   

The coarse-graining of the phase space, but in a different form than that described in the previous paragraph can be attributed to the 
approximate knowledge that we have  about the system, even at the (quasi-)classical level \cite{GellMH1, GellMH2}. 
In a physical situation there is always some
uncertainty, either about the dynamics or about the exact initial conditions of the system, or about both.  Such uncertainties are 
frequently encoded in dynamics as ``noise" or some other stochastic process through which the system interacts with its environment.
``Noise", or particular slowly varying background fields, can also be seen as encoding the collective effect of degrees of freedom in the system 
which although present may be considered of secondary importance  at the energy, time etc scale of 
interest. This is the spirit behind the Langevin approach in constructing kinetic equations \cite{vanKamp}.\\
 
In addition, since it is impossible to prepare a system with absolute accuracy  at some pre-determined state, we are inevitably led to 
consider not only a desired, or convenient, initial condition in our models, but  a set of initial conditions that are reasonably close to the desirable one
for the level of accuracy that we can tolerate in our predictions. Hence, one has to consider the evolution of sets of initial conditions
under the given dynamics, with or without  stochastic sources. This uncertainty is expressed by performing a periodic ``$\varepsilon$-fattening"
of the phase space evolution (orbit) of the system. After some judiciously chosen, for a particular model,  amount of time, one ``fattens" the 
Hamiltonian orbit, so that it initially appears to be like a tube. The question that arises is how such perturbations, assumed 
to be initially small, either in the initial conditions or through noise, affect the system under study. The initial hope that they would not 
affect the underlying system in any qualitatively significant  way was proved to be too naive in \cite{Smale1} (see also \cite {Smale2}), 
in the case of systems with phase space dimension $\mathrm{dim} \mathfrak{M} >2$. This instability directly questions the physical relevance, for Statistical 
Mechanics, of the single orbit analysis of any particular dynamical system, even if it were practically feasible. Hence one is forced to consider the 
behaviour of sets of orbits which are initially near each other. Then one uses the ergodic theorem to substitute averages over orbits 
with averages with respect to appropriate measures over the whole phase space.  
This is a reasonable choice, assumed to be true, as ergodic measures are precluded from having a ``complicated" phase space behaviour
such as possessing attracting sets etc. This is  direct implication of Birkhhoff's ergodic theorem  \cite{KatokHas}.    
Either way, and irrespective of the reason or the way that one chooses to perform phase space
coarse-graining, during such process some of the features of the underlying microscopic evolution are lost, a fact which is desirable 
if one wishes to capture the thermodynamic behaviour of the system with the half-dozen or so (at most) macroscopic variables, as is usually the case.\\              

The question that arises then is how to perform the coarse-graining of phase-space. The process appears, and largely is, ad hoc. 
But it is fundamental for the definition of any entropic functional. Due to its importance one may wish to make such a process a bit less less ad hoc by 
employing even partial knowledge about the underlying dynamics of the system.  The obvious choice is to assume that  the phase-space is divided
into cubical cells of side length $\sqrt{\hbar}$ each of which has obviously a volume  $\hbar^{n/2}$, if \ $\mathrm{dim}\mathfrak{M} = n$. \  
That typical cells in the coarse-graining process should be 
cubes is not only supported by the fact that geometrically they express ``independence" or due to their geometric simplicity, but also due to  
quantum nature of the underlying physical mechanisms. \\

Probably the simplest realisation of this underlying quantum nature is the emergence of the unit cube of side 
$\sqrt{\hbar}$ in the asymptotic expression for the spectrum of the Laplacian on $\mathfrak{M}$ which is provided by Weyl's asymptotic  formula \cite{ANPS}.
Weyl's asymptotic formula applies to a bounded domain $\Omega \subset \mathbf{R}^{n}$, but this is not a problem in our case, since the cubes that 
we use to coarse-grain  $\mathfrak{M}$ have such a small side that can be considered effectively flat,  to a first order approximation. Assume that 
such a domain  $\Omega$ has also a smooth boundary and we indicate by $\lambda_k, \ \ k=0, 1, \ldots, n, \ldots$ the eigenvalues of the Laplacian 
on functions $f: \Omega\rightarrow \mathbf{R}$ subject to the Dirichlet boundary condition $f|_{\partial\Omega} = 0$. Then the number $N(\Lambda)$ 
of such eigenvalues which are smaller than $\Lambda >0$ behaves asymptotically as    
\begin{equation}  
    N(\Lambda) \  \sim \  \frac{vol B_1 \cdot vol \Omega}{(2\pi\hbar)^\frac{n}{2}} \ \Lambda^\frac{n}{2},    \hspace{15mm}    \Lambda\rightarrow\infty  
\end{equation}
where we have used cubes of side length $\sqrt{\hbar}$. This  counts the number of quantum states of the Laplacian inside $\Omega$ which can also be 
re-interpreted as the number of quantum cubical cells inside $\Omega$, for macroscopic values of $\Lambda$ such as the ones needed in thermodynamics,
hence  $\Lambda\rightarrow \infty$. Such validity relies tacitly on the fact that fundamental kinetic terms are always quadratic and that long memory effects
that may give rise to non-Markovian evolutions described by anomalous kinetic terms are always an effective description arising due to the underlying statistics. \\    

Despite the above plausibility arguments, the choice of the fundamental cells to be cubes still remains somewhat arbitrary. It should also be considered as
still not acceptable as it ignores a central aspect of the Hamiltonian dynamics on phase space: its canonical transformation invariance, or in other words 
the existence of a symplectic structure on $\mathfrak{M}$. As will argue in the next Sections, choosing cubical cells for coarse-graining is probably 
the worst choice that someone could make in a metric sense, but probably the best in a measure-theoretical one: the best choice of a shape for the 
fundamental cells from the viewpoint of the Hamiltonian evolution would be (Euclidean) balls/ellipsoids instead of cubes. \\  

%%%%%%%%%%%%%%%%%%%%%%%%%%%%%%%%%%%%%%

\subsection{\normalsize  Riemannian aspects of phase space coarse-graining.}

An additional subtlety stems from the fact that the phase space on which the Hamiltonian evolution takes place is not usually $\mathbf{R}^n$ but some
Riemannian manifold $\mathfrak{M}$, with additional structure which we chose to overlook  in the previous Sections. Even though any Riemannian 
manifold can be $C^1$-differentiably embedded \cite{Nash1} (or even smoothly, i.e. $C^k, \ 3 \leq k \leq \infty$ embedded \cite{Nash2}) into some 
$\mathbf{R}^N$, for $N$ large enough,  an intrinsic description is sought after that would allow us not to worry about intrinsic vs the embedding 
features in the resulting geometric description. This is very much in the spirit of Geometry since the time of K.F. Gauss and was  implemented 
in General Relativity, for instance.  Riemannian manifolds are metrically almost Euclidean. Many of their metric properties can be 
expressed in terms  of their sectional curvature, which determines locally (second order deviation from ``flatness") the distances on $\mathfrak{M}$ 
\cite{Sakai, Gromov1, Gromov-book}.  
Among by-products of the sectional creature, the Ricci curvature determines the volumes of shapes lying in hyperplanes perpendicular to a 
given direction  on $\mathfrak{M}$, such as the direction of the Hamiltonian evolution. As a result of such curvatures, an initial shape will be distorted even if 
parallel transported along a curve. Hence if someone starts by partitioning the phase space into cubical cells, for the purposes of coarse graining, and wishes to 
follow the dynamics, the corresponding cells will become distorted cubes, i.e. $2n$-face polytopes, along any orbit of the 
Hamiltonian system. As long as the underlying dynamics remains invertible, then the number of faces of such polytopes will remain $2n$, even if the areas of 
the faces will no longer be equal to each other.  If, for whatever reason (such as taking the thermodynamic limit) the dynamics loses its invertibility 
\cite{KatokHas}, then such cells may acquire a larger or smaller number of faces.\\  
  
To summarise the discussion of this Section, coarse-graining and the curvature features of the phase space $\mathfrak{M}$ force us to consider not only 
cubes but more general polytopes as the basic cells of coarse-graining of phase-space $\mathfrak{M}$.
The need for such generalisation from cubes to polytopes becomes obvious, if one wishes to
incorporate in the formalism the effects of generalized products such as (12), (21) via their induced generalised concepts of independence  
which are expressed geometrically through their  induced ``unit" cubes such as (36), (37). In all this discussion so far,  we have (on purpose) ignored 
the dictates of the symplectic structure of $\mathfrak{M}$, which as will be seen in the next Section, point toward a very different, and largely incompatible, 
proposal on how to actually  perform such a coarse-graining.  \\

%%%%%%%%%%%%%%%%%%%%%%%%%%%%%%%%%%%%%%%%%%%%%%%%%%%%%%%%%%%%%%%%%%%%%%%%

                                                                                     \section{\large \ Symplectic basics: capacities and the role of ellipsoids.} 

                                                                                        \vspace{3mm}

In this Section, we provide some background on aspects of Symplectic Geometry/Topology that we need for our arguments, in an attempt to make the 
manuscript reasonably self-contained. Even though the concepts and facts that we  present are very well-known to a mathematical audience, some of them are 
very non-trivial  and have either only been proved relatively recently or they are still a subject of investigation. One might wish to consult some books, such as   
\cite{Arn, HoferZehn, McDSal, Polt, Schlenk, Zehn, Vit-Book}  or reviews \cite{Eliash, CHLS, McDuff}  to get a grasp of such matters that we can only very 
superficially touch upon here.\\ 

It may, for a moment,  be worth thinking about the role of the Hamiltonian approach to Mechanics. There are several, well-known advantages over 
the Lagrangian formulation (and vice-versa).  From our perspective, and for our purposes,  the Hamiltonian approach is more suitable
because it allows for more symmetries between its variables. By elevating the canonical coordinates \  $q^i, i=1,\ldots, n$ \ and the canonical 
momenta (not probabilities!)  \ $p_i, \ i=1,\ldots, n$ \ to equal status, the number of independent variables is doubled, hence there is greater possibility to detect and 
profitably use otherwise hidden symmetries or invariances. To make this easier to understand, we can start from simple discrete case: suppose we have  
been given one point. Then there is very little  in terms of operations and symmetries that one can detect, therefore very little latitude and 
substantial lack of direction in building, detecting or utilising such structures. Now consider a set whose elements are multiple copies of this point. 
Then one can easily start by determining its automorphism group and its algebraic properties, one can build discrete geometric structures such 
as graphs or simplices etc   and then by some from of reduction one can go back to the induced properties of such structures pertinent to one point.
This approach seems to have been appreciated first by E. Galois. The spirit of the Hamiltonian approach follows, to an extent, these lines. \\

%%%%%%%%%%%%%%%%%%%%%%%%%%%%%%%%%

\subsection{\normalsize Basics about symplectic vector spaces.}

Let $\mathcal{H}$ denote the Hamiltonian of a system (of many degrees of freedom, eventually). Hamilton's equations, as is well-known, are  
\begin{equation}
          \dot{q}^i \ = \ \frac{\partial \mathcal{H}}{\partial p_i}, \hspace{10mm} \dot{p}_i \ = \ - \frac{\partial \mathcal{H}}{\partial q^i}, \hspace{10mm} i=1, \ldots, n
\end{equation}
were the dot indicates differentiation with respect to the evolution parameter (``time").
Since the canonical coordinates and momenta are on equal footing in the Hamiltonian approach, we can put them side by side as coordinates of a vector  
\begin{equation}
       \xi \ = \ (q^1, \ldots, q^n, p_1, \ldots, p_n)
\end{equation}
and re-express Hamilton's equations as
\begin{equation} 
           \dot{\xi}^i \ = \ \omega^{ij} \ \frac{\partial \mathcal{H}}{\partial \xi^j}, \hspace{10mm} i,j,k = 1, \ldots, 2n 
\end{equation}
where the summation convention over repeated indices is assumed, and the matrix $\omega$ has elements \ $\omega_{ij}, \ \  i, j = 1, \ldots, n$ \ given by
\begin{equation}
      \omega_{ij} \ = \ \left( 
               \begin{array}{cc}
                     \mathbf{0}_n & -\mathbf{1}_n  \\
                     \mathbf{1}_n & \mathbf{0}_n 
               \end{array}   \right)
\end{equation}
where $\mathbf{0}_n$ and $\mathbf{1}_n$ stand for the null and the unit $n\times n $ matrices with real entries. Moreover, we see that 
\begin{equation}   
       \omega^T \ = \ -\omega, \hspace{15mm} \omega^{-1} \ = \ -\omega, \hspace{15mm} \omega^2 \ = \ -\mathbf{1}_{2n\times 2n} 
\end{equation}
These statements are abstracted in the definition of a real (finite dimensional) symplectic vector space $\mathfrak{V}$  
 which is a finite dimensional real vector space $\mathfrak{V}$, of even dimension, endowed with an antisymmetric and non-degenerate bilinear form $\omega$, namely
 \begin{equation}  
    \omega (X, Y) \ = \ - \omega (Y, X), \hspace{15mm} X,Y \in \mathfrak{V}
\end{equation}
and such that for any \ $X\neq 0 \in \mathfrak{V}$ \  there is \  $Y\in\mathfrak{V}$ \ such that 
\begin{equation}
   \omega (X, Y) \ \neq \ 0    
\end{equation}
The last equation is a non-degeneracy condition providing an isomorphism between $\mathfrak{V}$ and its dual $\mathfrak{V}^\ast$ by 
\begin{equation}
     X \ \longmapsto \  i_X\omega \ = \omega (X, \cdot)
\end{equation}
where $i_X$ denotes contraction of the symplectic form in the direction of the vector $X$. 
A different way to express the non-degeneracy of $\omega$ is by requiring that 
\begin{equation}
           \frac{\omega^n}{n !}
\end{equation}
where the \  $n !$ \ just fixes a normalisation, be a volume form on $\mathfrak{V}$, which is unique (up the normalisation).   
Following the standard algebraic practice, one defines $\mathfrak{W}$ to a be a symplectic subspace of $\mathfrak{V}$ if $\omega$ is non-degenerate 
on the linear subspace $\mathfrak{W}$. Obviously, the antisymmetry condition of $\omega$ is satisfied on $\mathfrak{W}$. \\

Even at this level, one can see that symplectic vector spaces are substantially different from spaces endowed with symmetric bilinear forms.
Requiring antisymmetry, instead of symmetry, of a bilinear form on such a space has proved to have profound consequences, some of which will be 
noted below. One such consequence is that the concepts of symplectic and Euclidean orthogonality are very different: Let $\mathfrak{U}$ be a linear 
subspace of the symplectic vector space $\mathfrak{V}$. Then the (symplectic) orthogonal complement $\mathfrak{U}^\perp $ of $\mathfrak{U}$ is 
defined by 
\begin{equation}       
    \mathfrak{U}^\perp \ = \ \{ X \in\mathfrak{V}: \omega (X,Y) = 0, \ \  \ \forall \ Y\in\mathfrak{U} \}  
\end{equation}
Unlike the case of Euclidean geometry, $\mathfrak{U}$ and $\mathfrak{U}^\perp$ need not be complementary subspaces, even though the 
non-degeneracy condition (44) implies that 
\begin{equation}
     \dim \mathfrak{U} + \dim \mathfrak{U}^\perp \ = \ \dim \mathfrak{V} 
\end{equation}  
On the one hand, if they are indeed complementary, namely if 
\begin{equation}
     \mathfrak{U} \oplus \mathfrak{U}^\perp \ = \ \mathfrak{V}     
\end{equation}
then one can prove this is equivalent to stating that $\mathfrak{U}$ is a symplectic subspace $\mathfrak{W}$ of $\mathfrak{V}$ which is also equivalent to 
stating that
\begin{equation} 
    \mathfrak{U} \cap \mathfrak{U}^\perp \ = \ \{ 0 \}
\end{equation}
On the other hand, one can observe that  that every vector $X\in\mathfrak{V}$ is orthogonal to itself due to the antisymmetry of the symplectic form (42). 
These relationships between $\mathfrak{U}$ and $\mathfrak{U}^\perp$ that are absent in Euclidean geometry can be generalized:  
$\mathfrak{U}$  is called isotropic if $\mathfrak{U} \subset \mathfrak{U}^\perp$, co-isotropic if $\mathfrak{U}^\perp \subset \mathfrak{U}$ and Lagrangian
if $\mathfrak{U}$ is both isotropic and co-isotropic, namely $\mathfrak{U} = \mathfrak{U}^\perp$. Clearly, in 2-dimensions, the lines passing by the 
origin are Lagrangian subspaces of the plane. The subspace of canonical coordinates and that of canonical momenta are Lagrangian subspaces
at a point of the phase-space  of a Hamiltonian system. \\

 All these definitions are ``strange" by Euclidean standards, 
as the Euclidean metric has been explicitly defined to exclude such occurrences. However the vanishing of a bilinear form may be more familiar
 in the context of Special, and General Relativity. If $\omega$ were a symmetric, rather than antisymmetric, bilinear form then the fact that 
\begin{equation}
      \omega(X, X) \ = \ 0, \hspace{15mm} X\in \mathfrak{V} 
\end{equation}
would define the light-like vectors $X\in\mathfrak{V}$. Since the usual $4-\dim $ Minkowski ``distance function"  
\begin{equation}
        ds^2 \ = \ - c^2 dt^2 + dx^2 + dy^2 + dz^2
\end{equation}
where $c$ indicates the speed of light, can be formally seen as arising from the $4-\dim $ Euclidean metric through the formal substitution  
\begin{equation}
       t  \longmapsto -it 
\end{equation}
one may be lead to suspect that there is an intimate relation between a Euclidean metric, an (almost) complex structure and the symplectic structure 
in a vector space $\mathfrak{V}$. The (almost) complex structure of $\mathfrak{V}$ can be defined as an anti-involution $\mathbf{J}$, namely 
$\mathbf{J}: \mathfrak{V} \rightarrow \mathfrak{V}$ such that 
\begin{equation}   
       \mathbf{J}^2 \  = \ - \mathbf{1}
\end{equation}
In more concrete terms and for $\mathfrak{V} = \mathbf{R}^{2n}$, with respect to a Cartesian base $\mathbf{J}$ has the antisymmetric form 
 \begin{equation}
      \mathbf{J} \ = \   \left(
           \begin{array}{cc}
                \mathbf{0}_n    &  \mathbf{1}_n \\
                -\mathbf{1}_n   &  \mathbf{0}_n  
           \end{array}     
                          \right)
\end{equation}
Then $\mathfrak{V}$ can be made into a complex vector space by defining, for $a,b\in\mathbf{R}$ and $X\in\mathfrak{V}$ 
\begin{equation}
     (a+ib) X \ = \ aX + b\mathbf{J}X
\end{equation} 
where the action of  $\mathbf{J}$ on $X$ has the effect of  a multiplication by $-i$.  It is no confidence that (42) and (56) have a similar form: 
indeed, if we indicate by $\langle\cdot , \cdot\rangle$ the Euclidean inner product on $\mathbf{R}^{2n}$ and for $X,Y \in\mathbf{R}^{2n}$  
we see that   
\begin{equation}
    \omega (X, Y) \ = \ \langle \mathbf{J}X, Y \rangle
\end{equation}
or, since $\mathbf{J}^2 \ = \ -\mathbf{1}_{2n \times 2n} $
\begin{equation}
     \omega(X, \mathbf{J}Y) \ = \ \langle X, Y \rangle
\end{equation}
The compatibility conditions (58), (59) have profound consequences in the case of manifolds to which we will turn in the next paragraphs. \\

Before that though, it may be worth mentioning another difference between the symplectic and the Euclidean cases: one can prove that 
in any symplectic vector space $\mathfrak{V}$ one can choose a ``symplectic basis" where the symplectic form will have essentially the 
same form as in (42), or to be more precise, one can pick a basis \ \ $e_i, \ f_j, \ \ i,j = 1, \ldots, n$ \ \ so that 
\begin{equation}   
   \omega (e_i, e_j) \ = \ 0, \hspace{15mm} \omega(f_i, f_j) \ = \ 0, \hspace{15mm} \omega (f_i, e_j) =  \delta_{ij}, \hspace{15mm}  i,j = 1, \ldots , n 
\end{equation}
From our experience with Hamiltonian dynamics, one can see that this is an abstraction of the fact that, locally, the symplectic form 
looks like (42) in each 2-plane made up of the canonical coordinate $q_i$ and its conjugate canonical momentum $p^i$ for $i=1, \ldots, n$. 
Hence (58) is the antisymmetric/symplectic analogue of the Gram-Schmidt diagonalization process of symmetric bilinear forms. We see that 
even though in the latter case there are numerous possibilities in this diagonalization process, in the symplectic case, all symplectic vector 
spaces are locally the same. This is behind Darboux's theorem and the lack of local symplectic invariants for the case of symplectic manifolds
in sharp contrast to the Riemannian case.\\

Let $\mathfrak{V}$ be a symplectic vector space endowed with the symplectic form $\omega$. A linear map $\varphi:\mathfrak{V} \rightarrow \mathfrak{V}$    
is called symplectic or canonical if it preserves the symplectic structure, namely if the pull-back form $\varphi^\ast \omega$ obeys
\begin{equation}        
    (\varphi^\ast \omega) (X,Y) \ = \ \omega (\varphi (X), \varphi (Y))
\end{equation}
As is clear from the terminology, symplectic maps are the linear canonical transformations of Hamiltonian Mechanics. These maps can be 
represented as matrices $\Phi$, and it turns out that they obey
\begin{equation}
        \det \Phi = 1
\end{equation}
whose geometric interpretation is that they are volume-preserving. This is the linear formulation of Liouville's theorem. 
The obvious question on whether there is any difference between volume preserving and symplectic maps, in the case of manifolds, will 
be discussed in the sequel. To complete the discussion of symplectic vector spaces, it turns out that if
$(\mathfrak{U}_1, \omega_1)$ and $(\mathfrak{U}_2, \omega_2)$ are two symplectic vector spaces of the same dimension, then there is a 
linear isomorphism $\varphi: \mathfrak{U}_1 \rightarrow \mathfrak{U}_2$ such that $\varphi^\ast\omega_2 \ = \ \omega_1$. Hence symplectic 
vector spaces of the same dimension are symplectically equivalent (indistiguishable from a symplectic viewpoint) as was also previously mentioned.   \\

%%%%%%%%%%%%%%%%%%%%%%%%%%%%%%%

\subsection{ \normalsize  About symplectic manifolds.}

To define symplectic manifolds $\mathfrak{M}$, one follows the same steps as in the Riemannian case, but substitutes the symplectic for the 
corresponding Euclidean   structures. Hence someone picks local patches (``charts") of symplectic (instead of Euclidean) vector spaces and 
glues them together using symplectic (rather than regular) diffeomorphisms. 
The details of such intuitively obvious, but non-trivial and cumbersome at  times, construction 
can be found in the references. It is worth mentioning at this point one consequence of the fact that we use symplectic diffeomorphisms to glue 
together patches of symplectic vector spaces: the symplectic form $\omega$ is postulated to be closed on $\mathfrak{M}$
\begin{equation}   
    d\omega = 0 
\end{equation}
where $d$ denotes exterior differentiation on the space of differential forms of $\mathfrak{M}$. 
One way to interpret the requirement (63) is to  use the canonical symplectic base (60) translated to the case of $\mathfrak{M}$.
A non-degenerate 2-form (hence anti-symmetric) $\omega$ is closed (63) if and only if at each point of $\mathfrak{M}$ 
there are coordinates $(q_1, \ldots, q_n, p_1,\ldots, p_n)$ such that 
\begin{equation}      
         \omega \ = \ \sum_{i=1}^n \ dq_i \wedge dp_i
\end{equation}
This theorem is due to G. Darboux. It expresses the fact that all symplectic manifolds are locally symplectically indistinguishable. Hence any non-trivial invariants  
of such manifolds will have to be global. Contrast this with the Riemannian case: in the Riemannian case there are plenty of local invariants which are 
encoded through the Riemann tensor at each point of $\mathfrak{M}$ and its multiple  covariant (properly symmetrized) derivatives and their contractions.
One can also see the lack of local structure of symplectic manifolds ``equivariantly": symplectic structures can be seen as the ``quotient" of a 
topological space locally homeomorphic to $\mathbf{R}^n$ under a set of ``symmetries" (actually re-paramentrizations, therefore more akin to gauge 
rather than global symmetries) the action of the group of symplectic diffeomorphisms. If the set of such symmetries is large enough, it is entirely possible 
that the resulting structure is unique: this is what actually happens in the case of symplectic manifolds, locally at least.    
The non-degeneracy condition (45) carried over to the case of a symplectic manifold $\mathfrak{M}$ can be seen as expressing, via the isomorphism (46), 
an isomorphism between the vector fields and the one forms of $\mathfrak{M}$, namely its tangent $T\mathfrak{M}$ and cotangent bundles 
$T^\ast \mathfrak{M}$. Consider a vector field $X\in T\mathfrak{M}$ and the Lie derivative along it, indicated by \cite{Sakai} $\mathcal{L}_X$, of $\omega$. 
Then according to Cartan's formula \cite{Sakai}, we find 
\begin{equation}
   \mathcal{L}_X \omega \ = \ d(i_X\omega ) + i_X d\omega
\end{equation}
If we assume that \ $\omega$ \ is closed \ ($d\omega = 0$), \ then
\begin{equation} 
     \mathcal{L}_X \omega \ = \ d(i_X\omega) 
\end{equation}
Due to the non-degeneracy condition(45), for any smooth function $f: \mathfrak{M}\rightarrow\mathbf{R}$ (``Hamiltonian")
there is a unique vector field $X_f: T\mathfrak{M} \rightarrow \mathbf{R}$ (``Hamiltonian vector field") such that 
\begin{equation}  
         i_{X_f} \ = \ df
\end{equation}   
Substituting (67) into (66) one gets that 
\begin{equation}
   \mathcal{L}_{X_f} \omega \ = \ 0
\end{equation} 
Therefore $\omega$ remains invariant under the flow generated by $X_f$. This is very desirable and natural from a physical viewpoint: 
in the case that $f = \mathcal{H}$, we would like the symplectic (canonical) form to remain invariant under the (``time") evolution of the system. 
Turning the argument around, we see that the invariance under evolution of the symplectic form is equivalent to requiring it to be closed, 
something which is usually assumed from the outset without further explanation.\\

 A second point arising from the above short argument is to
see that the local uniqueness of the symplectic structure, expressed through Darboux's theorem,  can indeed be seen from an equivariant 
viewpoint as previously suggested. The set of ``symmetries" of the symplectic structures is infinite dimensional since it is generated by the 
``Hamiltonian vector fields" $X_f$ corresponding to any smooth enough functions $f$. This is a typical situation in topological field theories, 
for instance, and it is quite extensively employed in field and string theories on models with enough supersymmetries etc. 
It may be worth comparing this to the Riemannian case in which the isometry group of any metric is finite dimensional, something that allows 
for local structure that is  able to differentiate between different Riemannian spaces.\\   

Consider  a system which is described by an autonomous Hamiltonian $\mathcal{H}$ and let $S_{\mathcal{H}_0}\cong S_0$
indicate the level set $\mathcal{H} = \mathcal{H}_0$ \ where $\mathcal{H}_0$ is a regular value of $\mathcal{H}$. As a result, its inverse image in 
$\mathfrak{M}$ is of codimension-1 (hypersurface).   
Let $X_\mathcal{H}$ be the Hamiltonian vector field corresponding to $\mathcal{H}$. Following (46) 
\begin{equation}
d\mathcal{H} (X) \ = \ \omega (X_\mathcal{H}, X) 
\end{equation} 
which gives 
\begin{equation}
       d\mathcal{H} (X_\mathcal{H}) \ = \  \omega (X_\mathcal{H}, X_\mathcal{H})  
\end{equation}
Hence at each point of $\mathfrak{M}$, $d\mathcal{H}$ is the kernel of the map $T\mathfrak{M} \rightarrow \mathbf{R}$. In other words, the 
Hamiltonian vector field $X_\mathcal{H}$ is tangent to, and therefore it preserves, the level sets $S_0$.
This result is  not totally unexpected if one looks at it from the viewpoint of the local compatibility between the simplectic and the almost complex 
structure expressed in the case of linear spaces in (58), (59). The role of $\mathbf{J}$ (56) is to generalise the complex unit $i$, hence its action 
can be seen to amount to a rotation by a right angle. The usual gradient vector field $X_\mathcal{H}$ is perpendicular to the level set 
$S_0$. Hence to compute the symplectic gradient, following (58), (59)  one has to would rotate this by an appropriate 
right angle thus making it tangential to the level set $S_0$.      
The above statements of this paragraph are a  formal way of expressing the fact that
the trajectory in phase space $\mathfrak{M}$ of an isolated system, evolves in the constant energy hyper-surface $S_0$, 
a realisation  which lies at the foundation of the micro-canonical approach in Statistical Mechanics. 
Since the Lie derivative $\mathcal{L}_X$ is a derivation, and given (68), we get 
\begin{equation}
    \mathcal{L}_{X_\mathcal{H}} \frac{\omega^n}{n!} \ = \ 0 
\end{equation}   
which is the familiar Liouville's theorem on the invariance of the phase space volume $\omega^n/n!$ under Hamiltonian  
flows. One can state more in this context: assume that the object of interest is not a particular Hamiltonian function $\mathcal{H}$, but 
instead the level set $S_0$ and the symplectic form $\omega$. Then if there are two Hamiltonian functions $\mathcal{H}_1$ and $\mathcal{H}_2$ 
 for which $S_0$ is a level set for  generally distinct values of $\mathcal{H}_1$ and $\mathcal{H}_2$, then these two Hamiltonian functions 
 will have the same trajectories on $S_0$.  \\

%%%%%%%%%%%%%%%%%%%%%%%%%%%%%%%%%%

\subsection{\normalsize The symplectic non-squeezing theorem.}

 We saw that symplectic geometry, in sharp contrast to the Euclidean/Riemannian case, is fundamentally 2-dimensional, something which 
 can also be justified in the following way.
Let $D_1$ be a disk in $\mathbf{R}^2$ and $D_2$ a subset of $\mathbf{R}^2$ diffeomorphic to $D_1$. If $D_1$ and $D_2$ have the same area 
then there is a symplectic diffeomorphism $\varphi: D_1 \rightarrow D_2$. This is due to J. Moser.  Hence in 2 dimensions, ``volume preserving" and 
``symplectic" are adjectives that can be used interchangeably. 
Therefore in 2-dimensions what distinguishes symplectic manifolds from each other is their total volume. 
The question that arose is how much of all these statements can carried over in higher dimensions. 
A step toward answering this question is the Gromov (-Eliashberg) alternative \cite{Gromov-PDR}: the group of symplectic diffeomorphisms of a 
$2n$-dimensional connected symplectic manifold $(\mathfrak{M}, \omega)$ is  $C^0$-closed in the group of all diffeomorphisms of $\mathfrak{M}$ 
or its $C^0$-closure is the group of volume preserving diffeomorphisms of
$(\mathfrak{M}, \omega)$. The fact that the former of these two alternatives is what actually occurs, was proved in the fundamental \cite{Gromov-Pseudo}. 
This result is also known as the ``symplectic non-squeezing theorem" or even as ``the principle of the symplectic camel": consider the standard symplectic 
space $(\mathbf{R}^{2n}, \omega)$  parametrized by $(x_1, \ldots, x_n, p_1, \ldots, p_n)$ where each $p_i$ is canonically conjugate to the corresponding 
$x_i$. Let $Z_i(R)$ indicate the cylinder of radius $R$ based on the symplectic 2-plane $(x_i, p_i)$, namely 
\begin{equation}
      Z_i (R) \ = \ \left\{ (x_1, \ldots, x_n, p_1,\ldots, p_n)\in \mathbf{R}^{2n} : x_i^2 + p_i^2 \leq R^2 \right\}
\end{equation}         
Then, if there is a symplectic diffeomorphism $\varphi: \mathbf{R}^{2n} \rightarrow \mathbf{R}^{2n}$ embedding the ball (33) into $Z_j (R)$ we must 
have $r\leq R$. It should be noticed here that the exact choice of the symplectic 2-plane does not matter. Someone could choose any cylinder based on 
another symplectic 2-plane $(x_j, p_j)$ and the result would still hold. However this conclusion does not hold if the cylinder is based on an isotropic 
2-plane $(x_i, x_j)$ or $(p_i, p_j)$ as a local rescaling, leaving all other coordinates unaffected, given by
\begin{equation}
     \varphi (x_i, x_j) \ = \ \left( \lambda^{-1} x_i, \lambda x_j \right),  \ \ \ \ \lambda\in \mathbf{R} \backslash  \{ 0 \} 
\end{equation}     
is a volume preserving transformation which is moreover symplectic and  can still embed the ball $B(r)$ into the cylinder 
$Z_i(R)$ for $\lambda \leq \frac{r}{R}$. In words, what the symplectic 
non-squeezing theorem states is that it is impossible to squeeze a ball inside a symplectic cylinder if the ball's radius is larger than the radius of the 
cross-section (base) of the cylinder. This rigidity does not apply for isotropic cylinders. The non-squeezing theorem can be seen as providing an 
obstruction for the existence of symplectic embeddings and it clearly shows that the ``symplectic" and ``volume preserving" classes 
of diffeomorphisms are not the same in dimension higher than 2, unlike the 2-dimensional case.  \\

The non-squeezing theorem can be seen as a counterpart to Liouville's theorem (71) which states that the symplectic volume of phase space remains 
invariant under a Hamiltonian (more generally: divergence-free) vector field. Liouville's theorem allows for the arbitrary change of shape of any subset 
of phase space $\mathfrak{M}$ under symplectic/canonical transformations generated by a Hamiltonian vector field. This arbitrary change of shape, 
and in particular, the fact that its  image under canonical transformations can become arbitrarily ``thin" in phase space, very much like ``oil in water"
is at the heart of Boltzmann's explanation of the macroscopic time-irreversibility of physical processes on the face of their time-reversible microscopic 
dynamics \cite{Bal-P, Bal-Book, Coh, Gallav}. The symplectic non-squeezing theorem states that such an arbitrary change of shape of a given phase space volume 
is  simply not possible under canonical transformations and explicitly provides a limitation.  Hence in higher dimensions ``symplectic" and 
``volume-preserving" are quite different terms. It is currently unknown what the effect(s), if any, would be due 
to the non-squeezing theorem in the description of a  Hamiltonian system of many degrees of freedom is.  Would this provide some constraints in 
applying Boltzmann's irreversibility argument with macroscopic consequences,  or the presence of the many degrees of freedom would ``wash out" 
such ``small-scale" features, as the Central Limit Theorem does for independently distributed random variables? In many ways, Boltzmann's 
irreversibility argument lies on the solid ground of Katok's lemma \cite{Katok, KatokHas} which goes as follows: Let $\mathfrak{U}_1, \mathfrak{U}_2$ 
be two bounded domains of equal volume in $\mathbf{R}^{2n}$ both of which are diffeomorphic to the ball $B(r)$. Indicate by $A \triangle B$ the 
symmetric difference between the sets $A$ and $B$. Then  for every $\epsilon >0$ there is a Hamitonian $\mathcal{H}$ and an evolution parameter 
(``time") t, so that 
\begin{equation}            
     vol \ (\varphi_t (\mathfrak{U}_1) \cap \mathfrak{U}_2) \ < \epsilon
\end{equation}
where $\varphi_t$ indicates the phase space flow generated by $\mathcal{H}$ calculated at time $t$. Due to this lemma, indeed any subset of $\mathfrak{M}$ 
can ``turn and twist" and become thin, overall, under canonical transformations. What cannot happen however, according to the 
symplectic non-squeezing theorem, is the projections of such a shape, as it evolves, along the symplectic 2-planes to become thinner than the original ones.
As stated above, it is still unknown what are the effects of such a limitation on  the projections of the initial shape along the canonical 2-planes. 
From the viewpoint of the present work, one can claim that the concept of entropy, for Hamiltonian systems of many degrees of freedom,  
can be seen as a manifestation of this underlying symplectic rigidity described by the symplectic non-squeezing theorem.\\        

It may probably be worth noticing, that during the 30 years that have elapsed since the formulation and first proof of the non-squeezing theorem, 
several proofs different from the original one have also appeared in the literature, none of which is elementary or 
even relatively simple. In the face of this and in order to get a better feeling on why this theorem is true,  one may wish to be less ambitious 
and try to present a relatively simple proof of the non-squeezing theorem in the case of linear symplectic diffeomorphisms as indicated, for instance, 
in \cite{deGos-Luef}. The intuitive advantage of such an approach is that such a proof involves concepts more familiar to physicists. 
We find it strange that 30 years have passed since the proof of the non-squeezing theorem, but its significance has not been widely appreciated in Physics. 
The most notable exception, in our opinion, is the work of M. de Gosson and his collaborators, who have been consistently emphasizing the interpretation 
and implications of the non-squeezing theorem, mainly for the physical cases of systems lying on the interface between Classical and Quantum Physics
 \cite{deGos-Luef, deGos1, deGos2, deGos3, deGos4, deGos5, deGos6, deGos7, deGos8, deGos-Book}. In this work, we rely considerably 
 on the concept of ``quantuum blob" \cite{deGos3, deGos-Luef, deGos7, deGos-Book} which will  be presented shortly.  \\

%%%%%%%%%%%%%%%%%%%%%%%%%%%%%%

\subsection{\normalsize  About symplectic capacities.}

As it befits a fundamental work, the contribution of the non-squeezing theorem was profound in actually establishing symplectic geometry/topology as a distinct 
field of mathematics rather than an interesting, but mostly, afterthought. The work of Gromov \cite{Gromov-Pseudo} is also credited for developing concepts, 
such as the $\mathbf{J}$-holomorphic curves that have proved to be enormously influential in a variety of contexts, some of which have significant overlap with 
developments in string/brane theories \cite{Taubes-Book}. For our purposes, the significance of the non-squeezing theorem lies in that it provides an explicit 
construction for a class  of (global) symplectic invariants called ``symplectic capacities" \cite{EkelHofer1, EkelHofer2, HZ-Cap, HoferZehn, Zehn} whose definition is 
the following: consider the class of symplectic manifolds $(\mathfrak{M}, \omega)$, possibly with a boundary, of dimension $2n$. A symplectic capacity
is a map $c: (\mathfrak{M}, \omega) \rightarrow \mathbf{R}_+ \cup +\infty$ with  $c$ having the  following properties       
\begin{itemize}
  \item {\sf Monotonicity:}  If there is a symplectic embedding \ \  $\varphi: (\mathfrak{M}, \omega ) \hookrightarrow (\mathfrak{N}, \omega^\prime )$  \ \ then 
                                     \begin{equation}    
                                           c(\mathfrak{M}, \omega ) \ \leq \ c(\mathfrak{N}, \omega^\prime ) 
                                     \end{equation}  
  \item{\sf Conformality:}  
                                     \begin{equation}
                             c(\mathfrak{M}, \lambda \omega) \ = \ |\lambda | c(\mathfrak{M}, \omega ), \hspace{10mm}   \lambda \in\mathbf{R}\backslash \{ 0 \} 
                                     \end{equation}
 \item{\sf Normalization}: 
                                    \begin{equation} 
                             c(B_r, \omega) \ = \ c(Z_r, \omega) \ = \ \pi r^2
                                    \end{equation}
            where $B_r$ is the  radius $r$ ball and $Z_r$ is the cylinder of radius base $r$ lying on a symplectic 2-plane, both in $\mathbf{R}^{2n}$
            endowed with its standard symplectic structure (64). 
\end{itemize}
It should be noticed that the conformality condition (76) can also be expressed, for $\mathfrak{U} \subset \mathfrak{M}$ and a fixed symplectic 
structure, as    
 \begin{itemize}
   \item {\sf Conformality:}
      \begin{equation}
              c(\lambda \mathfrak{U}) \ = \ \lambda^2 c(\mathfrak{U})
      \end{equation}
 \end{itemize}     
The normalisation condition (77) can also be relaxed  by just requiring
\begin{itemize}
     \item  \sf{Weak normalization:} 
          \begin{equation}
                   c (B_1) \  > \ 0 \hspace{10mm} {\rm and} \hspace{10mm}  c (Z_1) \ <  \ +\infty
           \end{equation}
\end{itemize}
It is a non-trivial fact  that the symplectic capacities are invariant under symplectic diffemorphisms. The converse is partly true: a differentiable map $\psi$  
not necessarily invertible, of  $(\mathbf{R}^{2n}, \omega)$ that leaves the capacities invariant is either symplectic or anti-symplectic, namely it satisfies 
\begin{equation}
      \psi^\ast (\omega) \ = \ \pm \omega
\end{equation}
The existence of symplectic capacities is not obvious at all, but it is guaranteed by the validity of the symplectic non-squeezing theorem. Conversely, 
the existence of symplectic capacities implies the non-squeezing theorem. Obviously
\begin{equation}
     c(\mathbf{R}^{2n}, \omega) = \infty 
\end{equation}
but for a symplectic cylinder $Z_r$, which is also an unbounded set in $\mathbf{R}^{2n}$, its symplectic capacity is bounded and given by (77).  
By contrast, for a cylinder $\tilde{Z}_r$ based on an isotropic 2-plane of $\mathbf{R}^{2n}$ the symplectic structure vanishes (by definition), therefore 
\begin{equation}  
      c(\tilde{Z}_r, \omega) \ = \ +\infty
\end{equation}
Based on the results of the non-squeezing theorem one can define the lower Gromov width (or just ``Gromov width", or symplectic radius) of a subset 
$\mathfrak{U} \subset \mathbf{R}^{2n}$ by
\begin{equation}
   c_{\min} (\mathfrak{U}) \ = \  \sup_\varphi \  \{ \pi r^2 : \varphi (B_r) \subset \mathfrak{U} \} 
\end{equation}
which means that the lower Gromov width is the maximum radius for which a ball of such a radius can be symplectically embedded through via $\varphi$
into $\mathfrak{U}$. Hence, if $c_{\min} (\mathfrak{U}) = r_0$, then $B_r$ cannot be symplectically embedded in $\mathfrak{U}$, if $r>r_0$.
By analogy,  the upper Gromov width (or cylindrical capacity) of $\mathfrak{V}\subset \mathbf{R}^{2n}$ is defined as   
\begin{equation}
    c_{\max} (\mathfrak{V}) \ = \ \inf_\psi \  \{  \pi R^2 : \psi (\mathfrak{V}) \subset Z_R \}
\end{equation}
where $Z_R$ is a symplectic cylinder of base radius $R$ lying on any symplectic 2-plane of $\mathbf{R}^{2n}$. This also means that the smallest radius of the 
symplectic cylinder inside which $\mathfrak{V}$ can be symplectically embedded via the symplectic map $\psi$ is $R$ and it is impossible to find a symplectic 
embedding of $\mathfrak{V}$ to a symplectic cylinder with a smaller radius than $R$ of its base. 
Given these two definitions, and based on the non-squeezing theorem, one sees that the lower Gromov width is the smallest possible capacity and the the 
upper Gromov width is the largest possible. Moreover, one can check that their convex combination 
\begin{equation}
   c_t \ = \ t \ c_{\max } \ + \ (1-t) \ c_{\min}, \hspace{15mm} t\in [0,1]
\end{equation}
is also a symplectic capacity. Hence there is an infinity of symplectic capacities on $\mathbf{R}^{2n}$.  
Despite this fact, constructing explicitly such  capacities has proved to be a  substantial challenge: to this date several such capacities have been constructed,  such 
as \cite{Gromov-Pseudo, EkelHofer1, EkelHofer2, Hofer,  HZ-Cap, CHLS, Hutch},  none of which is obvious to either construct or prove that they indeed obey the 
axioms (75)-(77) or (75), (76) and (79).  Without going into any details, we can indicatively mention that the Hofer-Zehnder capacity for compact, convex  
$\mathfrak{U} \subset \mathbf{R}^{2n}$ takes the form of an integral along the shortest  periodic orbit $\gamma_0 $ on the boundary of $\mathfrak{U}$ of the 
first Poincar\'{e} invariant $pdq$ familiar from Hamiltonian mechanics, namely
\begin{equation}
     c_{HZ} (\mathfrak{U} ) \ = \  \int_{\gamma_0 }  p_i \ dq_i
\end{equation} 
where the summation convention for \ $i=1, \ldots, n$ \  has been assumed. \\

%%%%%%%%%%%%%%%%%%%%%%%%%%%%%%%%%

\subsection{\normalsize Symplectic capacities of ellipsoids and the uncertainty principle.}

Calculating explicitly the symplectic capacities of manifolds, or even subsets of $\mathbf{R}^{2n}$ has proved to be a  difficult task. 
Among the very few cases for which an answer is known, the ellipsoids in $\mathbf{R}^{2n}$ stand out, because all symplectic capacities 
have the same value on them. This is straightforward for someone 
to see based on the symplectic non-squeezing theorem. Suppose that we have a real ellipsoid, the smallest two axes of which are of equal length $l$. 
Then a sphere of the  radius $l$ can be barely embedded in the ellipsoid, and the ellipsoid itself barely fits in a symplectic cylinder 
of the base radius $l$. Hence the upper and lower Gromov widths of this ellipsoid are $l$, so all its  symplectic capacities are proportional to $l^2$.\\
 
This argument can be made more concrete as follows \cite{deGos-Luef}. Let $A$ be a $2n \times 2n$ real positive-definite matrix, and $\mathbf{J}$ be as in (56).  
The eigenvalues of the matrix $\mathbf{J}A$ have the form \ $\pm i\lambda_k$ \ with $\lambda_k >0$ and are called the symplectic eigenvalues of $A$. 
The set $\lambda_k, \ k=1,\ldots, n$ is called the symplectic spectrum of $A$. 
Williamson's theorem states that there is a unique  element $B$ of the symplectic group $Sp(2n, \mathbf{R}^{2n})$ such that 
\begin{eqnarray} 
         B^t A B \ = \  \left(
                 \begin{array}{ll}
                       \Lambda_n & 0 \\   
                            0 & \Lambda_n                           
                \end{array}  \right)          
\end{eqnarray}
where the superscript $T$ stands for transposition and $\Lambda_n = \mathrm{diag} (\lambda_1, \ldots, \lambda_n)$. 
Parametrize $\mathbf{R}^{2n}$ by the row vector $z = (z_i), \ i=1, \ldots, 2n$ and consider the ellipsoid 
\begin{equation}
    \mathcal{E}: \ \ z^t A z \ \leq \ 1
\end{equation}
Then for any symplectic capacity $c$ one finds that 
\begin{equation}
  c(\mathcal{E}) = \ \frac{\pi}{\lambda_{\max }}
\end{equation}
where 
\begin{equation}
        \lambda_{\max } = \{ \max_k \ (\lambda_k), \ \ \ k=1,\ldots, n\} 
\end{equation}
Hence, from a Hamiltonian/symplectic viewpoint, the most appropriate/natural cells in which one should divide the phase space during coarse-graining are  
the ellipsoids (88). \\ 

It is a non-trivial fact \cite{deGos-Luef} about which we would not like to elaborate, 
that the Heisenberg uncertainty principle, or its generalisation, the Robertson-Schr\"{o}dinger inequality  
\begin{equation}
   (\Delta x_i)^2 (\Delta p_i)^2 \ \geq \ (\Delta (x_i, p_i))^2 + \frac{\hbar^2}{4}
\end{equation} 
where $\Delta (x_i, p_i)$ stands for the covariance matrix element, can be re-cast in terms of the symplectic capacities as 
\begin{equation}
     c( \mathcal{W} ) \ \geq \frac{\hbar}{2}
\end{equation} 
where $\mathcal{W}$ is the Wigner ellipsoid associated to the covariances and $c$ a symplectic capacity. This falls within the framework of the 
Wigner-Weyl approach to quantisation which is extensively used in the more mathematically rigorous or the quantum/classical interface treatments.     
In more familiar terms, one can see the content of the uncertainty principle as being expressed by the dictum that a function and its Fourier transform 
cannot be simultaneously sharply localized \cite{Fef, Fol-Sit}. Since we know that coherent states (Gaussians) represent minimum uncertainty states, 
and that they are mapped into Gaussians under the Fourier transform, let's assume
that a wave-function $\psi : \mathbf{R}^{2n} \rightarrow \mathbf{C}$ and its Fourier transform $\mathcal{F} [\psi]$ are bounded, in modulus,  
by such Gaussians, i.e.
\begin{equation}      
       | \psi (x) | \ \leq \ C \exp ( - \frac{1}{2\hbar} A x^2 ), \hspace{10mm}  | \mathcal{F} [\psi] (p) | \ \leq \ C \exp ( - \frac{1}{2\hbar} B p^2 )
\end{equation}
where $C$ is a constant and $A, B$ are real symmetric matrices. Then consider the phase-space ellipsoid 
$\mathcal{E}: \ Ax^2 + Bp^2 \ \leq \ \hbar$. The Robertson-Schr\"{o}dinger uncertainty principle can be expressed via a symplectic capacity $c$ by
\begin{equation}
    c(\mathcal{E}) \ \geq \ \frac{1}{2} \hbar
\end{equation}
These conditions imply that from a symplectic viewpoint, the appropriate choice of cell that should be  used for phase-space coarse-graining 
is an ellipsoid rather than a cube. This is also compatible with, if not necessarily dictated by,  Quantum Mechanics and provides a justification, in part,
for our approach an interpretations presented in the sequel.    \\

%%%%%%%%%%%%%%%%%%%%%%%%%%%%%%%%

\subsection{\normalsize Symplectic vs Riemannian features.}

It should be noticed that the symplectic capacities for an $n$-dimensional manifold $\mathfrak{M}$ are genuinely new invariants that cannot be deduced from its 
Riemannian volume \ $vol \ \mathfrak{M}$ \  by setting, for instance
\begin{equation}
    c (\mathfrak{M}) \ = \ ( vol \ \mathfrak{M})^\frac{2}{n} 
\end{equation}
as this would violate (77) for a symplectic cylinder, for instance.
On the other hand, we see that if the underlying manifold $\mathfrak{M}$ is compact, then it will have a finite volume hence 
\begin{equation} 
      c_{min} (\mathfrak{M}, \omega) < +\infty
\end{equation}
which is true for all compact manifolds. In the special case of 2 (real) dimensions, it is known that all symplectic capacities coincide with the area, as long as the 
manifold $\mathfrak{M}$ is connected and simply connected \cite{Sib, Jiang}. Hence in two dimensions the symplectic and volume-reserving geometries 
coincide.  But two dimensions are special in symplectic geometry.  There is a suspicion / conjecture that this statement may be partly true (``all symplectic 
capacities coincide") for convex bodies in $\mathbf{R}^{2n}$ but there seems to be neither a proof nor a counterexample needed to resolve it.  
A fundamental, generally still unresolved, question is whether there exist intermediate symplectic invariants \cite{CHLS}. The simple-sounding question 
about finding the necessary and sufficient conditions for an ellipsoid to be symplectically embeddable in another is still generally unresolved; 
an answer became only recently known in 4-dimensions  \cite{McDuff}. \\   

At this stage, there is an obvious question about the possible relation between Riemannian and symplectic invariants of a manifold $\mathfrak{M}$. As noticed 
previously, no such obvious relation exists for a power of the volume (95). But, the capacities $c$ are a symplectic way of measuring the size of a manifold. 
On top of that, in the phase space of a Hamiltonian system $\mathfrak{M}$  there is a Riemannian metric $\mathfrak{g}$, which is usually deduced from the 
quadratic term (``kinetic term") of the Hamiltonian. Therefore, one may  wish to compare the areas of (real) 2-dimensional sub-manifolds $\mathfrak{V} \subset 
\mathfrak{M}$ expressed via their embeddings $\varphi: \mathfrak{V} \rightarrow \mathfrak{M}$. 
Such an area is  computed symplectically via the pullback of the symplectic form 
\begin{equation}    
        A_{s} \ = \ \int_\mathfrak{V} \varphi^\ast (\omega)  
\end{equation}
or in a Riemannian manner by the area formula \cite{Fed}
\begin{equation} 
       A_R \ = \ \int_\mathfrak{V} \varphi^\ast (\mathfrak{g}) \ dvol_\mathfrak{M} 
\end{equation} 
where we use the volume on $\mathfrak{V}$ induced by the pullback metric given by the embedding $\varphi$. It turns out that 
(97), (98) are equal for pseudo-holomorphic curves $\mathfrak{V}$ \cite{Gromov-Pseudo, McDuff-Sal} which come about due to the extension  
of (58), (59) to almost complex target manifolds $\mathfrak{M}$, if endowed with a tame symplectic structure $\omega$. Such pseudo-holomorphic
curves turn out to be minimal surfaces in the Riemannian sense, hence can be considered  as the analogue of geodesics in symplectic geometry.
As strings move in space-time by ``sweeping out" surfaces that should be of stationary (``minimal") area with respect to variations of the area functional, 
according to the principle of ``least"/stationary action, the pseudo-holomorphic curves have been of great interest to string theory for the last two decades.\\    

Not much is generally known about the relation between symplectic and Riemannian invariants of manifolds. 
A prominent role in such a relation is furnished by the  conjecture of C. Viterbo \cite{Vit} which is formulated for convex domains in $\mathbf{R}^{2n}$.
Let $\mathfrak{U}$ be such a convex domain and $B_1$ be the Euclidean unit radius ball in $\mathbf{R}^{2n}$ (33). 
For any symplectic capacity $c$ and any such convex $\mathfrak{U}$ one has 
\begin{equation} 
      \frac{c(\mathfrak{U})}{c(B_1)} \ \leq \ \left( \frac{vol \ (\mathfrak{U})}{vol \ (B_1)} \right)^\frac{1}{n}
\end{equation}
where $vol$ denotes the Riemannian volume of the corresponding sets. One immediately sees that this conjecture follows the same philosophy  
as the failed attempt to relate the volume and the symplectic capacity (95). The big difference between (95) and (99), is that (99) makes a similar, in spirit, 
statement but expressed in relative terms, i.e. relative to the corresponding quantities for a ball. 
The meaning of this symplectic isoperimetric conjecture  is that among all convex 
domains $\mathfrak{U}$ in $\mathbf{R}^{2n}$ with a given volume, the Euclidean ball has the maximal symplectc capacity. 
Such a statement has a clearly isoperimetric flavour (``why is a droplet/bubble spherical"? or, among all shapes having a fixed volume, determine the one(s) 
with the least boundary/surface area). Viterbo's  conjecture (99) is not fully proved yet, although 
weaker versions of it have been proved, which rely on inserting a constant $A(n)$ on the right hand side of (99)     
\begin{equation}
     \frac{c(\mathfrak{U})}{c(B_1)} \ \leq \  A(n) \left( \frac{vol \ (\mathfrak{U})}{vol \ (B_1)} \right)^\frac{1}{n}
 \end{equation} 
Initially C. Viterbo proved (100) \cite{Vit} for $A(n) \sim n$, in particular $A(n) \sim 2n$ for symmetric convex domains with respect to the origin, 
and $A(n) \sim 32n$ for general convex domains. After that \cite{Art-Ostr}  improved the estimate to $A(n) \sim (\log n)^2$.       
The best known estimate, to our knowledge, was furnished by \cite{Art-Mil-Ostr} with $A(n)$ becoming an actual constant, i.e. 
independent of the dimension $n$ altogether, namely  $A(n) =  A_0$. It should be noticed that the assumption of $\mathfrak{U}$ being convex is essential:
indeed, star-shaped domains were constructed in \cite{Hermann} having an arbitrarily small volume but fixed capacity, which violates (99). The conjecture 
itself is true for ellipsoids and convex Reinhardt domains \cite{Hermann}.  
In the exact opposite direction of Viterbo's conjecture, namely in finding the worst possible symplectic capacities to volume ratios, 
one can see the symplectic cylinders are candidates, since this ratio is  zero in their case. \\

We see from the preceding analysis that ellipsoids are in some sense minimal and still are invariant under symplectic diffeomorphisms. 
Hence, from a purely Hamiltonian/symplectic  viewpoint it makes more sense to use as cells in coarse-graining the phase space $\mathfrak{M}$ 
ellipsoids rather than cubes. This is in sharp contrast to the Euclidean/Riemannian viewpoint of the previous Section for which, as we saw,  
cubical cells appear to be the most appropriate for such a coarse-graining process. Quantifying aspects of this mismatch between ellipsoids/spheres 
and cubes (or polytopes/convex bodies) to which we will ascribe the origin of entropy, is the topic of the next Section.\\            

%%%%%%%%%%%%%%%%%%%%%%%%%%%%%%%%%%%%%%%%%%%%%%%%%%%%%%%%%%%%%%%%%%%%%%%%%%%%%

                                                    \section{ \large \ Basic concepts and implications of convexity.}

                                                                       \vspace{3mm}

In this Section we will be using convexity exclusively in $\mathbf{R}^n$. This is not as big of a compromise as it may appear since all manifolds, 
symplectic or not, are locally isometric to $\mathbf{R}^n$ up to first order deviations. Curvature appears as a second order deviation from the 
Euclidean metric of $\mathbf{R}^n$. Hence, if one focuses on local properties of a manifold, understanding convexity of subsets of $\mathbf{R}^n$ 
already accomplishes quite a bit. Moreover, as was mentioned above, Nash's embedding theorems show that a generic manifold is not ``too 
different" from a Euclidean space since it can be isometrically embedded in a Euclidean space of high enough dimension. 
Working in $\mathbf{R}^n$ allows us to use its linear structure
to arrive in results that would not otherwise  be accessible. Since we are working with Hamiltonian systems of many degrees of freedom, we should be able 
to eventually consider the thermodynamic limit. Even though taking such a limit can be a non-trivial, model dependent,  process, it may not be 
unreasonable to assume, naively, that it is related to the limit $n\rightarrow \infty$. Hence we are interested in the behaviour of convex subsets of 
$\mathbf{R}^n$, for $n$ very large. This is the realm of the ``local theory of Banach spaces", or ``asymptotic geometric analysis", or ``asymptotic 
convex geometry". It turns out that there are highly non-trivial and unexpected / counter-intuitive results in this realm 
(such as the ``concentration of measure") which lies somewhere between linear algebra (for $n$:fixed, finite) and 
functional analysis (where one deals with infinite dimensional spaces).  
We would like to know about geometric structures that are, in a sense, ``typical", so they can encode results of importance for
Statistical Mechanics. The field of asymptotic convex geometry is highly developed. For our very limited purposes for this work, we have drawn, 
in various degrees, material  from the books \cite{Rock, Lind-Jafr, Mil-Sch, Pis, Benya, Schn, AAGM} and the reviews \cite{Ball, Versh, GM1, GM2} 
to which one can turn, as well as to the numerous other outstanding references,  for details and proofs  and authoritative and 
comprehensive expositions on these topics In the sequel, we will confine ourselves to concepts and results needed in the present work. \\
            
%%%%%%%%%%%%%%%%%%%%%%%%%%%%%%%%%            
            
\subsection{\normalsize  Convex bodies, polar and functional dualities.}

A convex set $\mathcal{K} \subset \mathbf{R}^{n}$ is a set such that 
\begin{equation}
     \forall \ x, y \in \mathcal{K}, \ \ \ \  tx +(1-t)y \in\mathcal{K}, \ \ \ \  t \in [0,1]
\end{equation} 
A convex body is a compact, convex subset of $\mathbf{R}^n$ having a non-empty interior. Hence balls, ellipsoids, cubes, polytopes etc. are convex 
bodies. A fundamental theorem in convexity is that Hahn-Banach separation theorem which implies that  each convex body $\mathcal{K}$ is an 
intersection of half-spaces and that at each point of the boundary $\partial\mathcal{K}$ of such a convex body there is at least one supporting hyperplane. 
Intuitively, this should be obvious. We will be mostly interested in convex bodies that are symmetric with respect to the origin of $\mathbf{R}^n$.
A convex body $\mathcal{K}$ is symmetric with respect to the origin of $\mathbf{R}^n$ if $x\in\mathcal{K}$ implies $-x\in\mathcal{K}$. Symmetric convex 
bodies are of great interest for the following reason:  
consider a finite dimensional normed space $\mathfrak{X}$. Then it is possible to choose a bijection $\mathfrak{X} \rightarrow\mathbf{R}^n$ such that 
$\mathfrak{X}$ can be identified with $(\mathbf{R}^n,  \| \cdot \|)$ for some norm $\| \cdot \|$. What we have in mind is either the usual  
Euclidean norm, or a ``generalised" norm induced by generalized products such as (12) or (21). The unit balls, centered at the origin,  
$B_1 (\mathfrak{X})$ of such norms are symmetric convex bodies, with respect to the origin. Conversely, for any symmetric convex body $\mathcal{K}$ 
one can assign canonically a corresponding (Minkowski) $\| \cdot \|_\mathcal{K}$ norm given by
\begin{equation}                
      \| x \|_\mathcal{K} \ = \ \min \{ s\geq 0 : \ \ x\in s\mathcal{K}\}
\end{equation}
whose unit ball is $B_1(\mathfrak{X}) \ = \ \mathcal{K}$. As examples, with the notation of (33), one can see that $B_1(l_2^n)$ is the Euclidean ball, 
$B_1(l_\infty ^n)$ is the usual cube (35) whose edge is the interval $[-1, +1]$ of length $2$ units, so it is symmetric with respect to the origin.
Another example, needed for our future reference, is the $n$-cross polytope, which is defined as the convex hull (97) of the endpoints of the unit orthonormal 
coordinate vectors  $e_i, \ i=1\ldots, n$ in Cartesian basis of $\mathbf{R}^n$. 
The metric having the cross polytope and the unit ball is called the Manhattan/taxicab metric, or more concretely, the cross-polytope is $B_1(l_1 ^n)$. 
The significance of the cross-polytope, for our purposes, stems from the fact that it is the polar set of the unit cube $B_1(l_\infty^n)$.                                                                     
 Consider a convex set $\mathcal{K} \subset \mathbf{R}^n$. Then its polar set $\mathcal{K}^\circ$ is defined by                                                                  
 \begin{equation}                                                                 
              \mathcal{K}^\circ \ = \ \{ b\in\mathbf{R}^n: \  \langle a,b \rangle \leq 1, \ \ \forall \ a\in\mathbf{R}^n \}                                                    
 \end{equation}                                                                 
 where $\langle\cdot, \cdot \rangle$ is the Euclidean inner product on $\mathbf{R}^n$.  What polarity does is to exchange the faces with the vertices. 
 By inspection, a cube in $\mathbf{R}^n$ has $2n$ faces and $2^n$ vertices and the converse is true for the $n$ cross-polytope.
 Upon a more careful examination, one can see that the polar of an $n$-cube is the $n$ cross-polytope.\\ 
 
There is an equivalent functional way of expressing the above polar facts, mainly because $\mathbf{R}^n$ is a linear space. 
 Let $\mathfrak{X}, \mathfrak{Y}$ be Banach spaces, whose corresponding norms even though different from each other, will still be indicated 
 by $\| \cdot \|$ for simplicity of notation. Consider a linear operator $T: \mathfrak{X} \rightarrow\mathfrak{Y}$. 
 Such an operator is bounded if there is some constant $C>0$ such that 
 \begin{equation}                                                                   
     \| Tx \| \ \leq \ C \|x \|, \ \ \ \forall \ x\in\mathfrak{X}                                                             
 \end{equation}                                                                 
Then the infimum of such numbers is the operator norm of $T$, indicated by $\| T \|$, so it  is defined by 
\begin{equation}                                                                    
      \| T \| \ = \ \sup_{x\in\mathfrak{X} \backslash \{ 0 \} } \frac{\| Tx \|}{\|x\| } \ = \ \sup_{\| x \| = 1} \| Tx \|                                                              
\end{equation}                          
 In the case that $\mathfrak{Y} = \mathbf{R}$ then $T$ is called a linear functional on $\mathfrak{X}$. 
 The space of continuous linear functionals of $\mathfrak{X}$ endowed with the operator norm is called the dual space 
 of $\mathfrak{X}$ and it is indicated by $\mathfrak{X}^\ast$. The dual space of a Banach space tuns out to be a Banach space too.
 As is well-known from linear algebra such a $T$ is  an isomorphism if it is a bijection and in addition both $\| T \|$ and $\| T^{-1} \|$ are bounded. 
 Such an isomorphism is an isometry if 
 \begin{equation}                            
           \| Tx \| \ = \ \| x \|, \ \ \ \ \ \forall \ x\in \mathfrak{X}               
 \end{equation}                         
 According to the Riesz representation theorem, every element $a \in \mathfrak{X}^\ast$ has the form $a\mapsto \langle a, b \rangle$                                     
 for some $b\in\mathbf{R}^n$. The unit ball of $\mathfrak{X}^\ast$ is therefore
 \begin{equation}                                           
     B_1 (\mathfrak{X}^\ast) \ = \ \{ b\in\mathbf{R}^n : \langle a,b \rangle \leq 1, \ \forall \ a \in B_1(\mathfrak{X}) \}                                        
 \end{equation}                                             
  which due to (103) can be rewritten as 
 \begin{equation}                                              
       B_1(\mathfrak{X}^\ast) \ = \  \left( B_1(\mathfrak{X}) \right)^\circ                                       
 \end{equation}                                             
 Hence the convex concept of polarity of symmetric convex bodies in $\mathbf{R}^n$  corresponds exactly to the functional 
 analytic concept of duality of $n$-dimensional normed spaces. Given  the well-known duality 
 \begin{equation}
         (l_1^n)^\ast \ = \ l_\infty^n 
 \end{equation}
 are see that  (108) is the generalisation of the polar relation
 \begin{equation}
      B_1 (l_1^n) \ = \ (B_1 (l_\infty ^n))^\circ
 \end{equation}
  indicated in the discussion preceeding (104). \\
  
 %%%%%%%%%%%%%%%%%%%%%%%%%%%%%%%%%%%%%%
 
 \subsection{ \normalsize  The Banach-Mazur distance.} 
 
 We would like to compare, quantitatively, the coarse-graining of phase space by cubes (and their  images 
 under generalised operations induced by non-additive entropies, which are convex polytopes) 
 and the balls/ellipsoids that should be the cells of coarse-graining of phase space 
 as dictated by its symplectic structure. This can be accomplished by introducing a distance $d$ in the set of all convex bodies. 
 To  define such a distance consider two symmetric convex bodies $\mathcal{K}_1$ and 
 $\mathcal{K}_2$  of $\mathbf{R}^n$. Then such a distance $d$ is given by 
 \begin{equation}                       
          d(\mathcal{K}_1, \mathcal{K}_2) \ = \ \inf \left\{ \alpha\beta :  \ \   \mathcal{K}_1 \subseteq \alpha\mathcal{K}_2, \ \ 
                     \mathcal{K}_2 \subseteq  \beta\mathcal{K}_1, \ \ \   \alpha>0,  \  \  \beta >0   \right\}             
 \end{equation}  
 What this essentially states is that the distance between $\mathcal{K}_1, \mathcal{K}_2$ is the smallest number $\delta$ such that 
 if $\mathcal{K}_1$ can be barely inscribed in $\mathcal{K}_2$, then $\delta \mathcal{K}_1$ can be barely circumscribed around $\mathcal{K}_2$ 
 and vice versa, with the role of $\mathcal{K}_1, \mathcal{K}_2$ interchanged. This definition relies on dilations  of the  symmetric convex 
 bodies $\mathcal{K}_1, \mathcal{K}_2$ and therefore it is, not surprisingly, multiplicative. To get an additive distance, one should take the 
 logarithm of such $d$. Moreover, and since physical theories use extensively metric concepts, it may be worthwhile comparing (111)  
 with the well-known Hausdorff distance as well as the numerous other distance functions that someone can come up with, many of which 
 are disucssed in  \cite{Gromov-book}.  \\
                        
 Consider the n-dimensional spaces $\mathfrak{X}_1, \mathfrak{X}_2$ whose unit balls are $\mathcal{K}_1, \mathcal{K}_2$ 
 respectively, as explained the previous subsection. Then, following (111)  their Banach-Mazur distance is defined as
 \begin{equation}                       
         d(\mathfrak{X}_1, \mathfrak{X}_2) \ = \ \inf \{ d(\mathcal{K}_1, T\mathcal{K}_2 ) : T\in GL(n, \mathbf{R}) \}
 \end{equation}
 where $T$ is an operator which is an element of the general linear (Lie) group $GL(n, \mathbf{R})$. Essentially what $T$ does is to 
 move around ``rigidly" $\mathcal{K}_1, \mathcal{K}_2$  until their mutual distance (111) becomes as small as possible, i.e. one of them ``barely 
 fitting" inside and outside the other, and vice versa. This can be expressed more succinctly by stating that $d$ is such that 
 $\mathcal{K}_1 \subseteq T(\mathcal{K}_2) \subseteq d\mathcal{K}_1$ for any $T\in GL(n, \mathbf{R})$. Since, as seen in the previous subsection, 
 there is an intimate link between convex geometry in $\mathbf{R}^n$ and  functional analysis, one can further translate the Banach-Mazur distance (112) 
 in the functional analytic language as
 \begin{equation}          
      d(\mathfrak{X}_1, \mathfrak{X}_2) \ = \ \min \{ \| T \| \cdot \| T^{-1} \|, \ \ \ \  T:\mathfrak{X}_1 \rightarrow \mathfrak{X}_2 \ \ \ \mathrm{isomorphism} \}
 \end{equation}
 which is a form  of the Banach-Mazur distance between normed spaces also frequently encountered.\\ 
 
 %%%%%%%%%%%%%%%%%%%%%%%%%%%%%%%%%%%%%%%%
 
 \subsection{ \normalsize  The Banach-Mazur distance between a sphere and a cube.}
 
  Consider now the cube $I_n = [-1, +1]^n$ in $\mathbf{R}^n$.   It is easy to see that one can inscribe in it a ball of radius 1 and can circumscribe 
  around it a ball of radius $\sqrt{n}$, for any $n$, and one cannot do better than that. Hence, it is intuitively obvious that the Banach-Mazur 
  distance between a cube and a ball is $\sqrt{n}$. Therefore as the dimension increases the cube looks less and less like a ball: the 
  vertices of the cube ``move" further and further away from the centre of the ball, assumed fixed, or conversely 
  the ball ``curves more" as $n$ increases and therefore it becomes smaller and smaller inside a fixed cube. 
  To make this a bit more precise, one can start form the well-known formula for the volume of the Euclidean unit-radius ball in $\mathbf{R}^n$ 
  \begin{equation}
         vol \ B_1 \ = \ \frac{\pi^\frac{n}{2}}{\Gamma (\frac{n}{2} + 1)}
  \end{equation}
  where $\Gamma$ stands for the (Euler) gamma function. Using Stirling's approximation 
   \begin{equation} 
            \Gamma \left( \frac{n}{2} +1 \right) \ \sim \ \sqrt{2\pi} \ e^{-\frac{n}{2}} \left( \frac{n}{2} \right)^\frac{n+1}{2} 
   \end{equation}  
    provides the estimate 
   \begin{equation}
         vol \ B_1 \ \sim \ \left( \sqrt{\frac{2\pi e}{n}}\right)^n
   \end{equation}
   Therefore one can see that due to the curvature of the surface of the ball, the ball of unit radius has a volume that approaches zero very fast as 
   $n\rightarrow\infty$. This is in sharp contrast with the case of the cube $I_n$ whose volume increases as $n\rightarrow\infty$. 
   Hence it is expected that the distance between the cube and the ball of unit radius will increase as $n$ increases. Not only that, but such a distance
   is as large as possible for the ball and the cube, a direct outcome of F. John's theorem. \\
   
   Based on the above and from a metric viewpoint, as was also previously remarked, coarse-graining the phase space 
   $\mathfrak{M}$ with balls instead of cubes (or vice versa) should provide the greatest possible discrepancies than the coarse-graining of $\mathfrak{M}$ 
   with any other regular polytopes. But entropy, and this is probably clearest in the definition of the Kolmogorov-Sinai case \cite{KatokHas}, involves 
   considering the supremum  over all possible phase space partitions. Since coarse-graining involves a piece-wise (for each cell) constant  probability 
   density $\rho_{cg}$, phase space measures are constant multiples, per cell, of their phase space volumes. As a result, 
   substituting small enough   cubic cells for balls (and vice versa) in coarse-graining will provide the maximal possible measure discrepancy, 
   namely the maximal  possible phase-space volume loss, which is what $\mathcal{S}_{BGS}$ has been designed to capture. \\     
    
   It should be noticed that the polar duality is of fundamental importance in asymptotic convex geometry hence it may be desirable to compare pairs of
   polar dual polytopes rather than single convex bodies. In this respect the Euclidean ball is unique in that it is the only is self-dual 
   polytope under polarity. If it is pairs of polar dual polytopes that are of fundamental interest, the one should also consider the 
   Banach-Mazur distance between  $B_1 (l_2^n)$  and the cross-polytope $B_1 (l_1^n)$. It should also be intuitively obvious that such a 
  distance is $n$ and this is as bad as things can get. Intuitively the cross-polytope and the cube are the ``pointiest" of all convex bodies hence their 
  distance from the ``roundest" of all of them, which is the ball, should be maximal.  \\  
    
  For our purposes the ball, or more generally ellipsoids, are a manifestation 
  of the symplectic structure of $\mathbf{R}^n$ and the cube encodes geometrically the concept of  independence, as seen in the previous sections. 
  Since however we would like to allow for  generalised concepts of ``independence" induced by entropies such as $\mathcal{S}_q$, $\mathcal{S}_\kappa$ etc. 
  whose induced ``cubes" are, in general, symmetric polytopes/convex bodies what we would like to ask is what is the Banach-Mazur distance between 
  convex bodes and ellipsoids. And since actually calculating the Banach-Mazur distance between convex bodies has proved to be quite hard, in 
  general, one can settle either by asking for upper bounds for that distance or for asymptotic estimates as $n\rightarrow\infty$ (the ``thermodynamic limit").              
  That there  is a unique ellipsoid of maximal volume that can barely fit inside a convex body is guaranteed by F. John's theorem where conditions for 
  this maximal ellipsoid to actually be the Euclidean sphere are also spelled out.\\
  
   Another result, intuitively plausible, if not obvious, is that a reasons why a sphere and a cube are so different is that the cube has too few faces. What would 
   happen if one allowed for a symmetric convex body with far more faces? 
  Then the answer is that its distance form the sphere should decrease. And this is actually what is happening but the crucial matter, for our purposes, 
  is that the increase has to be exponential in terms of the dimension. More precisely, consider a symmetric convex body 
  $\mathcal{K} \subseteq \mathbf{R}^n$ such that $d(\mathcal{K}, B_1(l_2^n)) = d$. Then $\mathcal{K}$ must have at least \ $\exp (\frac{n}{2d^2})$ \  faces.        
  Since the distance between the cube and the ball is $\sqrt{n}$,  a cube in $\mathbf{R}^m$ has almost spherical sections of dimension 
  $\log m$. A substantial generalisation of this fact,  Dvoretzky's theorem,  we will be used in Section 6.\\
  
  Having stated that the distance between a cube an Euclidean ball is the maximum possible, we may turn to a 
  desirable consequence already embedded in the definition of entropy \cite{Bal-P, Bal-Book, Lesne}. Entropy, like any thermodynamic quantity,  
  involves ignoring a lot of the details of the underlying dynamical processes as noticed earlier in this subsection. 
  All such details are contained in the phase space evolution of the 
  underlying dynamical system. Hence skipping many of these details, for small enough cells in a coarse-graining of phase-space, 
  is tantamount to glossing over or even ignoring a substantial volume of the cells, if we think that such an omission will not make, statistically, any substantial 
  difference. Since $\mathcal{S}_{BGS}$ describes systems having a simple phase space evolution according to Birkhoff's ergodic theorem \cite{KatokHas}, 
  hence the micro-canonical density $\rho$ is uniform on a constant energy hypersurface of the isolated system, we can simplify our considerations and use 
  balls instead of cubes, as the distance between them is the largest possible, for coarse-graining. 
  By doing that, we have to ignore most of the volume of these cubes, thus simplifying the description, without 
  losing much, due to the uniformity of the micro-canonical measure and its proportionality to volume within each cell. \\
  
  This procedure is effective due to the  fact that the majority of the volume of the cube $I_n$ is close to  its vertices, something that can 
  be justified as follows: place the cube's centre of symmetry  at the origin of $\mathbf{R}^n$ and express its volume in spherical coordinates. Let 
  let $\theta$ collectively express the angular coordinates, parametrising the unit sphere $S^{n-1}$  and $r(\theta)$ be the radial coordinate. Then    
  \begin{equation}
          vol \ (I_n) \ = \ vol (B_1)  \int_{S^{n-1}} \  r^n(\theta) \ d\Sigma (\theta) 
  \end{equation} 
 where $d\Sigma$ expresses the infinitesimal area element of the surface of $I_n$.  
 Since $I_n$ is a cube of side length 2, its volume is $2^n$. Therefore  
  \begin{equation}
         \int_{S^{n-1}} \  r^n(\theta) \ d\Sigma (\theta) \ = \ \frac{2^n}{vol (B_1)} 
   \end{equation}
  which, after using (112) gives approximately 
  \begin{equation}
        \int_{S^{n-1}} \  r^n(\theta) \ d\Sigma (\theta) \ \sim \  \left( \frac{2n}{\pi e} \right)^\frac{n}{2}
  \end{equation}
  This corresponds to a cube of average radius about \ $\sqrt{2n \slash \pi e}$. \ Given that the distance between the vertices of the cube and its centre is 
  $\sqrt{n}$, and that centre of each face is $1$ unit away from its centre, we conclude from (119) that most of the volume of the cube is closer to its vertices 
  than to the centre of its faces. Hence substituting the cubes $I_n$ by Euclidean balls inscribed in them, hence of unit radius, omits most of the volume of the 
  cube, which however is rather innocuous, as can be inferred from the above discussion.\\
       
  We should be very careful however, since such an argument may not necessarily apply for systems described by $\mathcal{S}_q, \mathcal{S}_\kappa$ 
  or any of the other non-additive entropies. Such systems will have more complicated phase space behaviour, potentially attracting sets and the like \cite{Ruelle},
  which a regular measure $\rho$ may not be able to adequately describe. As the case of the Sinai-Ruelle-Bowen \cite{Young} measures indicates, 
  one may have to use measures some of the marginals of which may be nowhere absolutely continuous with respect to the phase space volume, 
  thus invalidating the above arguments and complicating substantially the coarse-graining process.  \\            
  
  %%%%%%%%%%%%%%%%%%%%%%%%%%%%%%%%%%%%%%%%%%%
  
  \subsection{ \normalsize  Unit Euclidean balls and Gaussians.}
  
  There is another reason for considering, and quantifying, the discrepancy between balls/ellipsoids and cubes/convex bodies/polytopes. Consider the, 
  arguably, simplest system of many degrees of freedom: the classical (non-relativistic) ideal gas of $N$ identical particles of mass $2$ units which is  
  placed inside an isolated cubical box of side length $L$.  The phase space of this system factorizes as $L^{3N} \times \mathbf{R}^{3N}$. 
  The Hamiltonian is 
  \begin{equation}    
              \mathcal{H} \ = \ \sum_{i=1}^N p ^2
  \end{equation}    
  Given that the system is isolated, its total energy is conserved. Set it equal to unit for simplicity. Hence the momentum space reduces from 
  $\mathbf{R}^{3N}$ to the unit sphere $S^{3N-1}$. The thermodynamic limit corresponds to taking $N\rightarrow\infty$ which gives rise to the 
  Maxwell distribution, which is Gaussian with respect tot the molecular speeds.  
  Probabilistically, this is a simple manifestation of the Central Limit Theorem. This result, interpreted geometrically, shows that a unit radius sphere 
  of high dimension ($N \rightarrow \infty$) should be an excellent approximation to the Gaussian distribution. \\
  
  That this is indeed true had been stressed by \'{E}. Borel and later by P. L\'{e}vy. 
  More recently, it has been emphasized in the work of V.D. Milman and M. Gromov.  The argument can be 
  made more precise as follows \cite{Ball}: consider the Euclidean ball $B^n_R$ of radius $R$ in $\mathbf{R}^n$. Due to the homogeneity of volume, 
  its volume is 
  \begin{equation}    
         vol \ B^n_R \ = \ R^n \ vol \ B_1
  \end{equation}                     
 where $B_1$ is given by (114).   
 Assume that $vol B^n_R = 1$. Hence 
 \begin{equation}  
        R \ = \ ( vol \ B_1 ) ^{-\frac{1}{n}} 
 \end{equation} 
 Hence a section of this ball of codimension 1 passing through the origin, will be an $n-1$ dimensional ball $B^{n-1}_R$ of radius $R$ 
 whose volume is according to (118) and (119)
 \begin{equation}
     vol \  B^{n-1}_1 R^{n-1} \ = \ vol \ B^{n-1}_1 \ (vol \ B^n_1) ^{-\frac{n-1}{n}} 
 \end{equation} 
 After using Stirling's approximation (115), we see that for large $n$ this section has a volume approximately equal to $\sqrt{e}$. 
 Consider now an $n-1$ dimensional section of this ball at a distance $r$ from the ball's center. Its radius will be  
 $(R^2-r^2)^\frac{1}{2}$. As a result its volume will be, for large $n$, approximately given by 
 \begin{equation}
       \sqrt{e} \left( \frac{(R^2 - r^2)^\frac{1}{2}}{R} \right)^{n-1} \ = \ \sqrt{e} \left( 1 - \frac{r^2}{R^2} \right)^\frac{n-1}{2} 
 \end{equation}   
Following (116) we can see that the a Euclidean ball of volume 1, for large $n$, has a radius of about
\begin{equation}
               R \sim \ \sqrt{\frac{n}{2\pi e}}
\end{equation}   
 Substituting (125) in (124)  gives for the volume of the spherical section at distance $r$ from the origin 
\begin{equation}
   \sqrt{e} \left( 1- \frac{2\pi e r^2}{n} \right)^\frac{n-1}{2} \ \sim \ \sqrt{e} \ \exp (-\pi e r^2) 
\end{equation}   
 So, the projection of the unit volume of the ball in a spherical section of co-dimension 1, which is at a distance $r$ from the ball's centre, for 
 large $n$, is  almost a Gaussian distribution of variance $\frac{1}{2\pi e}$. What we have done is a geometric re-formutation and ``derivation" 
 of the Central Limit Theorem. \\
 
 What we also observe in the above derivation is that most of the volume of the ball concentrates in a lower dimensional section passing through its center.
 This observation  turns out to be  independent of the validity of the Central Limit Theorem, 
 even if this is not obvious from the above arguments.
 Counter-intuitive as it may be, it is frequently encountered in convex asymptotic geometry, i.e. 
 in  Banach spaces as their dimension increases to infinity. Extensions for the cases of Riemannian manifolds with a Ricci curvature bounded from 
 below  or in terms of the behavior of the lowest non-trivial eigenvalue the spectrum of their Laplacian also exist. 
 The same can be stated for smooth metric measure spaces under additional assumptions 
 and generalizations of the definition of convexity.  This underlying behavior has been called "concentration of measure" \cite{Mil1, Mil2, GroMil1, 
 GroMil2, Mil-Sch, Mil3, Gromov-book, Ledoux, AAGM} and was the main avenue for V.D. Milman proving Dvoretzky's theorem which we will come to 
 in the next Section.\\
 
 %%%%%%%%%%%%%%%%%%%%%%%%%%%%%%%%%%%
 
 \subsection{ \normalsize  About ``quantum blobs".}
 
 In the previous Sections we saw that there is a substantial discrepancy between the coarse-graining approach that rely on cubes which are the 
 outcome of  independence/simplicity arguments and balls/ellipsoids induced from symplectic capacities. Ultimately one would like to have a ``natural way", 
 to the extent possible, to decide on such coarse-graining. The existence of the fundamental constant $\hbar$ arising in Quantum Physics, partially helps
 provide a scale for such a phase-space coarse-graining, but the exact shape of the fundamental cell employed is still a matter of choice, as previously has been 
 pointed out. Since balls/ellipsoids
 minimise all symplectic capacities and their high dimensional sections are Gaussian, as seen in the previous subsection, one may be willing to use them
 as the fundamental cells for phase-space coarse-graining. 
 This would be favourably supported by Quantum Mechanics where the minimum uncertainty, hence ``as precise as possible", 
 wave-functions for quadratic potentials, which are the lowest order approximations to any ``generic" analytic potentials, are Gaussians. See also the 
 discussion in subsection 4.5.  In this subsection 
 we rely on the work of M. de Gosson and collaborators \cite{deGos-Luef, deGos1, deGos2, deGos3, deGos4, deGos5, deGos6, deGos7, deGos8, 
 deGos-Book} where many more details on simiar topics can be found.\\ 
      
 ``Quantum blobs" are, very much like Gaussians, minimum uncertainty sets whose size is measured using symplectic capacities instead of volumes. 
 Their advantage, as opposed to cubic cells, is that they remain invariant under canonical transformations, hence they preserve the Hamiltonian structure 
 of the dynamical system.  To be more specific, we work in $\mathbf{R}^{2n}$, and define a quantum blob $\mathcal{Q}^{2n}(x_0)$ to be the image of the 
 Euclidean ball  $B_{\sqrt{\hbar}} (x_0) \subset \mathbf{R}^{2n}$ under a canonical/symplectic transformation. Using the results of subsection (4.5)
 we see that for a quantum blob 
 \begin{equation}
      c(\mathcal{Q}^{2n}(x_0)) \ = \ \frac{1}{2} \hbar 
 \end{equation}
 which has the same order of magnitude as the symplectic capacity of a cube. By contrast, since  symplectic transformations are volume-preserving, 
 the volume of a quantum blob, following (118), is  
 \begin{equation}
        vol \ (\mathcal{Q}^{2n} (x_0)) \ = \ \frac{h^n}{ n! \  2^n}  
 \end{equation}
 As  result, this volume is $n! 2^n$ times smaller than that of a cube. Since we are interested in the thermodynamic limit $n\rightarrow\infty$, the 
 quantum blob has far (``infinitely" ) smaller volume than that of a cube, despite the fact that its sections along all symplectic 2-planes are of 
 comparable area.\\     
 
 The above considerations of cubes versus ellipsoids as fundamental cells of phase-space coarse-graining raise another question. If the 
 two-dimensional sections of these two candidates for fundamental cells are of almost the same area but their volumes are so different, is there any 
 intermediate situation? Given that quantum blobs  have a substantially smaller volume than the corresponding cubes and that inside each cube, 
 or polytope, there is a maximal volume ellipsoid (F. John's theorem), is there any intermediate dimensional section of the cube which is close to being 
 a ball/ellipsoid?  \\

 %%%%%%%%%%%%%%%%%%%%%%%%%%%%%%%%%%%%%%%%%%%%%%%%%%%%%%%%%%%%%%%%%%%%%%%%%%%%  
   
                                    \section{ \large  \ Dvoretzky's theorem, Dvoretzky dimension and dualities.}  
   
                                                                         \vspace{3mm}

 In this Section we present Dvoretzky's theorem and the Dvoretzky dimension, and its implications for the definition of entropy. 
 We do not claim that we can actually predict the functional form of entropy to be used in each occasion, as would be desirable, if feasible, in our opinion. 
 However we can at least 
 see that the choices of $\mathcal{S}_{BGS}$ and $\mathcal{S}_q$ are plausible, asymptotically, for the case of the ``thermodynamic limit" 
 $(n\rightarrow\infty)$ \ seen through a perspective/viewpoint induced by Dvoretzky's theorem and its implications. 
 The choices of the appropriate deformation fields such as $\mathbf{R}_q$ and $\mathbf{R}_\kappa$ contribute in determining the shape of the fundamental 
 cells in which the phase space $\mathfrak{M}$ of the Hamiltonian system should be divided in a coarse-graining process. 
 However such a choice, if it is also followed by the other axioms 
 of Shannon / Khintchin or Abe / Santos etc amounts to the choice of the entropic functional employed in any particular situation. 
 Still the  viewpoint that we follow in this Section may be useful in seeing the entropy under a different light, which may allow inferences  that may not be easily 
 accessible otherwise, such as the origin of dualities in non-additive entropies etc. \\  
       
%%%%%%%%%%%%%%%%%%%%%%%%%%%%%%%%%%%%%%%%   
     
\subsection{ \normalsize  Dvoretzky's theorem, Dvoretzky's dimension and entropies.}   
   
   The above considerations of cubes versus ellipsoids as fundamental cells of phase-space coarse-graining raise another question. If the 
 two-dimensional sections of these two candidates for fundamental cells of phase-space coarse-graining are of almost the same 2-dim ``area" but 
 their volumes are so different, is there any  intermediate situation? Given that quantum blobs  have a substantially smaller volume than the 
 corresponding cubes and that inside each cube or convex body/polytope there is a maximal volume ellipsoid (F. John's theorem), is there any 
 intermediate dimensional section of the cube which is close to being a ball/ellipsoid? We saw in subsection 5.3 that for a symmetric convex body to be close 
 to a ball, it must have exponentially many faces. Equivalently, a cube in $\mathbf{R}^n$ has   almost spherical sections of dimension at least \ $\log n$. 
 Hence, if the dimension of a section of a cube is larger than $\log n$ then it can be reasonably close to a ball, therefore the discrepancy 
 between phase space coarse-graining by cubes and by balls can be seen as  non-significant. On the other hand, as seen in subsection 5.3,  
 if such a  section has dimension much larger than $\log n$, then the cube and the ball will be substantially different from each other in their volumes, 
 hence a coarse-graining procedure would give quite different results for these two fundamental cells. 
 Since cubes and balls are as distinct from each other as possible as measured by their Banach-Mazur distance, 
 getting the same coarse-graining results for both of them can be seen as relatively re-assuring that we will get the same results by using as 
 fundamental cell of phase space any other symmetric convex body (polytope).\\
 
 From a physical viewpoint, one can interpret this way of thinking to mean that $\log n$ as $n\rightarrow\infty$ is, asymptotically, the optimal order of magnitude 
 dimension that one can use in order  for the coarse-graining results of phase space to be virtually independent of the exact shape of the fundamental cell. 
 Hence the statistically important characteristics of the underlying dynamical system would be preserved, geometrically, but there would still be a reduction 
 in the complexity as measured by the number of the effective degrees of freedom of the system. But this is exactly what the entropy was designed to capture.
 Of course, the entropy is associated to a measure, which may not be just the volume. However, if one assumes the validity of the ergodic hypothesis, 
 the micro-canoncal density is uniform on the constant energy hyper surface under consideration, as was also previously pointed out (subsection 5.3). 
 Hence all the arguments pertaining to the micro-canonical measure are reduced to the corresponding ones about volumes.    
 Hence, in the most conventional sense, the entropy should have the form of a natural logarithm of the accessible phase space, in accordance with 
 Boltzmann's, Gibbs', Shannon's etc  ideas.\\
 
 One wonders on whether anything similar can be stated about non-extensive Statistical Mechanics and the non-additive entropies such as $\mathcal{S}_q$ 
 and $\mathcal{S}_\kappa$. After all, the underling geometric structures are almost the same, the only difference being that cubes will have to 
 be replaced by their images under (14), (23) and their generalisations, generalised shapes which are  symmetric convex polytopes. 
 The underlying symplectic structure, by contrast, remains the same as Hamilton's equations are still assumed to be applicable for such systems (see also the 
 related comment in subsection 3.3 after eq. (38)). Hence the question that arises, in analogy with the cube, is whether there is any dimension beyond which 
 a symmetric convex body (polytope) has almost spherical sections. There is an additional level of difficulty however: it is not clear that the systems described by 
 $\mathcal{S}_q$ or $\mathcal{S}_\kappa$ are ergodic. On the contrary, it has been conjectured that these non-additive entropies describe exactly non-ergodic 
 systems (see also the last paragraph of subsection 5.3). 
 Even though any ``reasonable" measure admits a decomposition into ergodic components \cite{KatokHas}, this is not enough to justify the 
 reduction of the phase-space measure to just the volume. To proceed, we assume that each phase space cell is so small that such a reduction is possible.     
 This can occur, for instance, when the measure density variation is slow, when compared to the spatial extent of each cell.\\ 
  
  To make the question more precise and general, assume that $X$ is an $n$-dimensional Banach space whose unit ball expresses the generalized independence 
  induced by some non-additive entropic functional.  Then, is there a subspace $E$ of $X$  of dimension $k(\epsilon, n)$ such that $d(E, l^k_2) \leq 1+\epsilon$, 
  for  $\epsilon >0$ ?   In this expression $d$ stands for the Banach-Mazur distance between $E$ and the $k$-dimensional Euclidean (Hilbert) space $l^k_2$. 
  Slightly more geometrically, one can start from the symmetric convex body $K$ in $\mathbf{R}^n$ which expresses the afore-mentioned independence.
  Does there exist a section $K\cap E$ of $K$ by a subspace $E$ of dimension $k(\epsilon, n)$ so that if $\mathcal{E}$ is an ellipsoid, one has the inclusion
  $\mathcal{E} \subseteq K\cap E \subseteq (1+\epsilon ) \mathcal{E}$? It should be noticed that the above questions are asymptotic, in the sense that $n$ is large,     
  namely $n\rightarrow\infty$.\ The answer to these equivalent questions is affirmative as was proved in 1961 by A. Dvoretzky \cite{Mil-Sch, AAGM, Ball, Mil1, Mil2, 
  GM1, GM2} a result which is one of the cornerstones of Geometric Functional Analysis / Asymptotic Convex Geometry. The result is extremely counter-intuitive
  as someone can easily see by trying to imagine a section of a cube which is almost spherical. Such a result is possible exactly because the dimension $n$ is 
  assumed to be quite large. \\
  
  As it befits a fundamental result, there are different re-formulations in slightly different contexts under the general title of 
  ``Dvoretzky's theorem" and several alternative proofs. One of the still unsettled questions is the optimal form of the Dvoretzky dimension $k(\epsilon, n)$ in the 
  above statements. The best known estimate is 
  \begin{equation}          
        k(\epsilon, n) \  \geq \ c\epsilon^2 \log n, \hspace{10mm} \epsilon \in (0,1)                             
  \end{equation}                                           
   The case of our interest (see also the discussion in the next subsection)  
   involves the $n$-dimensional Banach spaces endowed with the norms (30), (31) which were indicated as \ \ $l^n_p, \ 1\leq p \leq \infty$.                                                               
  It is a non-trivial result that the corresponding Dvoretzky dimensions $k(l^n_p)$ are given asymptotically $n\rightarrow\infty$ by
  \begin{equation} 
     k(l^n_p) \ \sim \   
                     \left\{
            \begin{array}{ll}
                  n, & 1\leq p \leq 2\\
                  p \ n^\frac{2}{p}, & 2 < p < \infty\\
                  \log n, & p=\infty\\     
            \end{array}         
                     \right.
  \end{equation}
  We observe that when $p>2$ then we have a power-law behaviour. If the Dvoretzky dimension of a cube $k(l^n_\infty)$ is responsible for giving rise to the 
  logarithmic form of the entropy as we previously suggested, then one can see  that power-law entropic forms  
  such as $\mathcal{S}_q$ and $\mathcal{S}_\kappa$ may be seen, at a first glance, as arising from the Dvoretzky dimension of $l^n_p, \ p >2$. 
  Even though this would not be exactly correct, in the face of the comments of the next subsection, at least it may be considered as suggestive, 
  and indicative of yet not fully appreciated underlying structures. We observe in (130) the remarkable property 
  that $k(l^n_p) \sim n, \ 1\leq p \leq 2$. Pushed to its limit, and substituting $q$ for $p$ in (130), if the above discussion is pertinent for $\mathcal{S}_q$,  
  then the assumed range of entropic indices $q \in (0, 1)$ would cover all possible cases for the leading power-law behaviour of the entropic functional. 
  If the conclusions of \cite{BL, LB} prove to be correct, the range $q\in (0,1)$  in $\mathcal{S}_q$ is the only acceptable one  for the entropic index related to a 
  Hamiltonian system of many degrees of freedom. Similar things can be probably stated about the entropic parameter $\kappa$ and the entropic 
  functional  $\mathcal{S}_\kappa$. \\
  
  Potentially existing sub-leading terms in the non-additive power-law functionals cannot be detected by the asymptotic form (130), so far as we can see.
  It is probably exactly such sub-leading terms that determine the differences between functionals such as $\mathcal{S}_q$ and $\mathcal{S}_\kappa$
  and therefore these terms may turn out to able to distinguish which one of these functionals  is more appropriate for describing which system.       
  Still, referring to more general systems described by the above two, or any other, entropic functionals of 
  power-law form,  we are not quite certain about how to interpret the significance, if any, of the ``phase transition" in the Dvoretzky dimension of $l^n_p$ 
  exhibited between $p \in [1,2]$ and $p \in (2,\infty)$.\\
  
  %%%%%%%%%%%%%%%%%%%%%%%%%%%%%%%%%%%%%%%
  
  \subsection{ \normalsize  On the ($\epsilon, \tau$) entropy and the use of  \ $l^n_p$ \  spaces.}
  
  It may be of interest to elaborate a bit upon the dynamical origin and role of $\epsilon$ appearing in Dvoretzky's theorem. This parameter expresses, as was 
  previously discussed in subsection 3.3, our fundamental inability to follow the evolution of the underlying dynamical system with absolute precision. One way
  to quantify this, is through a modification of the Kolmogorov-Sinai entropy \cite{KatokHas} such as the one presented in \cite{GW} called $\epsilon$-entropy:
  in the case of the Kolmogorov-Sinai entropy, one introduces partitions/cells  of the phase space of size $\epsilon$ over which one defines 
  $\mathcal{S}_{BGS}$. Eventually, one takes the size of the partition $\epsilon \rightarrow 0$. By contrast in the $\epsilon$-entropy the size of the fundamental 
  cell remains finite, but all other steps as the same as in the case of the Kolmogorov-Sinai entropy construction. One can refer to \cite{GW} for the advantages 
  and difficulties that such a definition implies. What is important, for our purposes however, is that in this definition the finite, even if small, phase-space 
  ``resolution" $\epsilon$ enters explicitly the definition entropy. Hence this parameter will invariably appear, explicitly or implicitly, in the 
  composition properties such as  (8), (9) or (19) and therefore $\mathbf{R}, \mathbf{R}_q$ or $\mathbf{R}_\kappa$ respectively.  
  Since one defines the fundamental polytopes of phase space coarse-graining, which express geometrically ``independence", via these algebraic operations and 
  structures, the shape of these cells/polytopes will invariably contain a dependence on $\epsilon$. Therefore there will be some finite uncertainty about     
  the exact shape of the convex body that one should use in Dvoretzky's theorem, based on the above physical arguments. This uncertainty is quantified and 
  appears in Dvoretzky's theorem as the upper bound requirement of the Banach-Mazur distance 
  in the theorem and explicitly in the Dvoretzky dimension $\kappa (\epsilon, n)$ as its dependence on $\epsilon$. 
  In this paragraph, we have used $\epsilon$ twice, in two quite different contexts. This is clearly a substantial abuse of notation: 
  even though one may possibly expect to find a relation between the indeterminacies expressed by $\epsilon$ in these two contexts, 
  one should not assume that they are equal, let alone the same. \\
  
  A second point that may be clarifying, is the extensive use of the \ $l^n_p, \ 1\leq p \leq \infty$, \ Banach spaces  in arguments (subsection 3.2 and forth) 
  in this work. One reason for such use is that a lot of their features are reasonably well-understood, when compared to general Banach spaces. This however 
  does not make them physically more relevant, just formally more tractable. Based on $\mathcal{S}_{BGS}$ and $\mathcal{S}_q$, $\mathcal{S}_\kappa$ it 
  appears   that the concept of independence as expressed by cubes, namely the unit ball of  $l^n_\infty$, should be sufficient for our purposes. 
  After all, the field isomorphisms (14), (23) are  invertible and distance non-decreasing. So they preserve the number of vertices of these cubes. All 
  that they do is to uniformly expand the sides of the fundamental cubical cells of the phase 
  space partition used in coarse-graining.  In addition, any legitimate cube should have clearly defined vertices. However, all $l^n_p, \ 1 < p < \infty$ have unit 
  spheres that are everywhere differentiable, hence they do not possess any clearly defined (point) vertices. \\    
       
  To address the last concern, one should refer to the ideas leading to the use of the ($\epsilon, \tau$) entropy in the first paragraph of this subsection. 
  In realistic models, even classical ones, there is always some uncertainty associated to the scale of phase-space coarse-graining $\epsilon$. This should be
  reflected on the composition property of the pertinent entropic functional. 
  This indeterminacy, in turn, makes the concept of independence become a bit ``vague":  the  
  corresponding cubes do not have well-defined vertices and faces, but rather areas of small but finite ``thickness" as faces, and areas of 
  small spatial extent / ``radius" as vertices. Hence one should not be able to distinguish between ``cubes" in $l^n_p$ for, let's say,  
  \ $p=1$ \ and for \ $p=1+\delta$, \ where \ $0 < \delta \ll 1$. 
  A second reason in favour of using nn only the spaces $l^n_p$ but also more general Banach spaces in employing Dvoretzky's theorem for physically 
   relevant cases is that we want to have a formalism that is flexible enough to accommodate,  many families of entropic functionals.
  If the price that one pays for such a flexibility is small, then one is willing to go along with a slightly more elaborate, but far more general, formalism to 
  accomplish these ends. Consider, for instance, models of highly anisotropic systems, of as practical as materials possessing layered structures \cite{BBH}, or as 
   exotic and conjectural as of anisotropic (Ho\v{r}ava-Lifshitz \cite{Hor} etc.) gravity. Then it may not be entirely unreasonable to propose a direction-depedent 
  entropic form for such systems, as long as there is a relatively clear separation of the dynamics and scales in the different directions. This would elevate 
  the non-extensive parameters of entropic functionals such as $\mathcal{S}_q, \mathcal{S}_\kappa$ into vectors. If this is true, then the 
  corresponding unit cells, expressing independence, in phase-space coarse-graining will be anisotropic convex bodies which however can still be 
  accommodated by the convexity formalism presented here. To this date though, there has not been any compelling theoretical
  reason to introduce any such vector-valued entropic index functionals, so far as we know. \\       
  
%%%%%%%%%%%%%%%%%%%%%%%%%%%%%%%%%%%%%%%%%

\subsection{ \normalsize  Polarity, Mahler's conjecture and symplectic rigidity.}

 An issue that may be worth discussing  is that of polar duality. From a geometric as well as analytical viewpoint, polarity has turned out to be of considerable 
 significance, since the earliest days of Euclidean geometry. The functional analytic analogy of  the polar of a convex body $\mathcal{K}^\circ$ with the 
 dual space $\mathfrak{X}^\ast$ presented in subsection 5.1 has important and far-reaching consequences.  One of them is that the Banach-Mazur distance        
 remains invariant under such a duality, namely for any normed spaces $\mathfrak{X}, \mathfrak{Y}$, the Banach-Mazur distance obeys 
 \begin{equation}
       d(\mathfrak{X}, \mathfrak{Y}) \ = \ d(\mathfrak{X}^\ast, \mathfrak{Y}^\ast) 
 \end{equation}
An immediate implication is that since, according to John's theorem
\begin{equation} 
    d(l^n_2, l^n_\infty) \ = \  \sqrt{n}
\end{equation}
and because $(l^n_1)^\ast = l^n_\infty$ and the Hilbert space $l^n_2$ is self-dual (the Euclidean ball is the polar of itself), one also has
\begin{equation}
    d(l^n_2, l^n_1) \ = \ \sqrt{n}
\end{equation}
Since the different $l^n_p$ via their unit balls express different ways of defining independence, induced by the various non-additive entropies, one also needs
the extension of the above to
\begin{equation}  
     d(l^n_p, l^n_q) \ = \ n^{\frac{1}{p}-\frac{1}{q}}
\end{equation}
where either $1\leq p \leq q \leq 2$ \ or \ $2\leq p \leq q \leq \infty$. The remaining option, namely  $1 \leq p \leq 2 < q \leq \infty$ gives only bounds for the 
 Banach-Mazur distance as
 \begin{equation}
      C_1 n^\beta \ \leq \ d(l^n_p, l^n_q) \ \leq  \ C_2 n^\beta,   \hspace{10mm} \beta \ = \  \max \ \left\{ \frac{1}{p}-\frac{1}{2}, \ \ \frac{1}{2} - \frac{1}{q} \right\} 
 \end{equation} 
with $C_1, C_2$ being positive constants. Going back to Dvoretzky's theorem, the Figiel-Lindenstrauss-Milman theorem provides for the Dvoretzky dimension 
$k$ of a Banach space  $\mathfrak{X}$ and its dual $\mathfrak{X}^\ast$ the lower bound
\begin{equation}
      k(\mathfrak{X}) \ k(\mathfrak{X}^\ast) \ \geq \ \frac{Cn^2}{d(\mathfrak{X}, l^n_2)^2}
\end{equation}
which since \ $d(\mathfrak{X}, l^n_2) \leq \sqrt{n}$ \ gives that 
\begin{equation}
     k(\mathfrak{X}) \ k(\mathfrak{X}^\ast) \ \geq \ C'n  
\end{equation}
where $C$ and $C'$ are positive constants. As a result, for any such Banach space $\mathfrak{X}$, one has that either \ $k(\mathfrak{X}) \geq C\sqrt{n}$ \ 
or \ $k(\mathfrak{X}^\ast ) \geq C\sqrt{n}$, \ a result that turns out to be sharp.\\

We refer to dualities in this work, not only because they play an important role in Convex Geometry, but also because they may be of 
importance for the case of non-additive entropies. It has been surmised from some data, for $\mathcal{S}_q$ for instance, that some systems seem to be 
invariant under the entropic parameter changes   
\begin{equation}
    q \ \longmapsto \ 2-q, \hspace{15mm} q \ \longmapsto \ \frac{1}{q}
\end{equation}
Whether this actually happens, and if so what is the origin of such invariances is still an open question. The transformations (138) 
are the generators of M\"{o}bius transformations for $q\in\mathbf{C}$. Clearly the case $q=1$ in (138) which corresponds to $\mathcal{S}_{BGS}$ 
remains invariant under such transformations, hence issues that can be raised for $q\neq 1$ pertaining to (138), are undetectable for $\mathcal{S}_{BGS}$.    
Convex polarity of the unit balls and the corresponding Banach space duality may somehow be related to (138) in a currently not understood manner.  
However a pattern that starts emerging that the above considerations are suggestive of,  is that it may be worthwhile  to analyse in parallel, and compare to each 
other, features of systems whose entropic indices are connected by some duality transformation.  Assuming that  such systems are described by different values 
of the entropic parameter of the same single-parameter family of functionals such as $\mathcal{S}_q$ or $\mathcal{S}_\kappa$, it may be worth investigating
what features of such systems that are common, or in some, still vague,  sense ``opposite"/``complementary".\\

To push this viewpoint a little bit further, it may be worth examining concurrently from both a convex and a symplectic geometric viewpoints properties of the 
unit balls of dual to each other finite dimensional  Banach spaces $\mathfrak{X}$ endowed with the norm $\| \cdot \|$ and $\mathfrak{X}^\ast$ endowed with 
the dual norm $\| \cdot \|^\ast$. To this end, one may consider examining convex and symplectic geometric properties of properties of $\mathfrak{X} \times 
\mathfrak{X}^\ast$. A straightforward observation is that this vector space has a canonical symplectic structure: assume that $X, Y \in \mathfrak{X}$ 
and that $X^\ast, Y^\ast \in \mathfrak{X}^\ast$ are their respective duals. Then the canonical simplectic structure 
$\omega$ on $\mathfrak{X} \times \mathfrak{X}^\ast$ is defined by        
\begin{equation}
     \omega ( (X,X^\ast), (Y, Y^\ast)) \ = \ X^\ast (Y) - Y^\ast (X)
\end{equation}
and the corresponding Liouville form of $\mathfrak{X} \times \mathfrak{X}^\ast$ is, of course, $\omega^n/n!$ . \ We saw during all this work t
hat the Euclidean ball and  the cube are, in some sense, as different from each other as possible, and that even though the former behaves quite well under 
symplectic transformations the same is not true for the latter. Therefore, it may come as a complete surprise that for the case of the cube 
$I_n \subset \mathbf{R}^n$ and its polar, the cross-polytope, $I_n^\circ \subset \mathbf{R}^n$, the interior of $I_n \times I_n^\circ$ turns out to be 
symplectically diffeomorphic to the interior of the  Euclidean ball in $\mathbf{R}^{2n}$ of the same volume \cite{Schlenk}. This again shows the 
unexpected features of symplectic geometry where flabbiness and rigidity can be found in totally unexpected places.\\

On the geometric side, a question in the spirit of the isoperimetric problem 
\cite{Gromov-book, Mil-Sch, Chavel} which was posed by Mahler (ca. 1939) was to find upper and lower bounds for the Liouville volume 
of $B_1\times B_1^\circ$ where $B_1, B_1^\circ$ stand for the unit balls of $\mathfrak{X}$ and $\mathfrak{X}^\ast$ respectively. 
This volume $vol (B_1\times B_1^\circ)$, often called ``Mahler volume", is invariant under linear invertible transformations.  The upper bound 
was determined in 2 and 3 dimensions in \cite{Blaschke}, and generalised in any dimension \cite{Santalo},  if and only if 
$\mathfrak{X}$ is the Euclidean space (Blaschke-Santal\'{o} inequality). The equality was proved in \cite{Petty}. Mahler conjectured that the lower bound is 
$4^n / n!$ and would be sharp. This lower bound would clearly apply for the pair of the cube its polar, the cross-polytope. Mahler himself verified this 
conjecture in 2 dimensions but in higher dimensions the conjecture remains unproven. What has been proved though is the conjecture up to multiplicative factor
whose best value known today is given in \cite{Kuper}. In an interesting recent development, \cite{Art-Kar-Ost, Ostrover} assumed the validity of the Viterbo 
conjecture  (eqs.(99), (100) and the discussion around them), and proved that the Hofer-Zehnder capacity for a symmetric convex body $\mathcal{K}$ and its polar 
$\mathcal{K}^\circ$ is
\begin{equation}        
      c (\mathcal{K} \times \mathcal{K}^\circ ) \ = \ 4
\end{equation} 
which, in turn, showed that the Viterbo conjecture implies the validity of the Mahler conjecture. \\

This subsection used some  symplectic and convex geometric facts and conjectures to suggest that it may be formally fruitful for someone to look at the same time 
at pairs of systems, rather than single systems,  described by entropies, belonging to the same single-parameter family but having harmonically conjugate indices 
(which represent geometrically polarity and 
Banach space duality). It remains to be seen whether this approach may provide some insights into the nature of such systems as well as about the possible 
invariances and their origin, of non-additive entropies such as $\mathcal{S}_q$ under ``duality" transformations such as (138). \\    

%%%%%%%%%%%%%%%%%%%%%%%%%%%%%%%%%%%%%%%%%%%%%%%%%%%%%%%%%%%%%%%%%%%%%%%%%%%%%

                                                          \section{  \large  \ Conclusions and discussion.}

                                                                                          \vspace{3mm}

In this work we presented the view that the source of entropy can be ascribed to two mutually exclusive ways of performing phase space coarse-graining.  
for Hamiltonian systems with many degrees of freedom. The underlying Euclidean/Riemannian structure favours cells that are cubical. By contrast, the 
symplectic structure favours ellipsoids. We discussed ways to measure the discrepancy of these two disjoint  approaches and also gave estimates, via 
Dvoretzky's dimension of minimal dimensions spaces in which they give almost the same results. So, it is in some sense, as if there is  a set of variables 
present at the microscopic level that reflects the number of variables that one sees as the outcome of a statistical analysis, i.e. in thermodynamics. We cannot 
quite dare claim that these variables are the same, or even more so, that there is a ``phase-space thermodynamic" behaviour. It just appears from Dvoretzky's 
theorem  that at the microscopic level one can infer a number of variables that are of the same order of magnitude as the ones needed for a macroscopic descriptio 
of the system.  Further investigation in this direction may be of some interest.   Moreover, we saw some preliminary formal indications about the suspected presence 
and about the origin of dualities of non-additive entropies via polarity. \\      

One could certainly expand this work in both the symplectic and the convex geometric directions, if deemed necessary. 
The symplectic capacities are still not very well understood objects. 
In case this looks too removed from the modelling of physical systems, it may be worth mentioning that there is  an elaborate and flourishing research 
area on the interface between symplectic geometry and string theory. Even though the goals and approaches in this area may appear substantially 
different from the ones of Statistical Mechanics, some  general ideas and technical approaches  especially pertaining to dualities \cite{Polch, HKKP, Ram}  
may be profitably adapted in the present context. 
After all, quantum perturbative string theory, like any quantum theory, has a statistical interpretation and explicitly uses methods of statistical mechanics relying on 
$\mathcal{S}_{BGS}$. To what extent one may wish to consider other functionals in such a statistical approach is unclear. However, given string 
theory's origins in dual resonance models that were  eventually superseded by QCD which is asymptotically free and has a  Wilson loop formulation 
\cite{Polyakov},  shows us that low energy correlations become dominant, a feature of systems that non-additive entropies such as $\mathcal{S}_q$ 
claim to describe. Hence it may be worth looking into string theory from an non-additive entropy viewpoint. There is more than just pure speculation on this front: 
using the phenomenological asymptotic bootstrap approach of Hagedorn for 
strong interactions, some recent results suggest an important role that $\mathcal{S}_q$ may play in this regime. Such phenomenological approaches
relying partly on $\mathcal{S}_q$,  seem to be, most importantly,  in accordance with existing experimental data \cite{UBB1, UBB2, Dep, WW1, MCD, WWCT}.\\

One can also use several well-known results of convex geometric analysis, such as the Bourgain distortion 
theorem or the Johnson-Lindenstrauss flattening lemma etc. to expand upon the 
results that just used Dvoretzky's theorem and the associated dimension \cite{Lind-Jafr, Pis, Benya, AAGM, GM1, GM2}. 
Whether such results can be generalized and can lead to interesting conclusions pertinent to non-additive entropies is not clear in our mind. 
However, the thought of using a physical idea, such as a non-additive entropic functional, to potentially help prove a purely geometric conjecture 
such as that of Mahler, is probably too enticing to not motivate someone to look carefully and further develop the symplectic/asymptotic convex 
point of view.\\   
 
In closing, and from a formalistic viewpoint, 
one could not avoid mentioning a trend toward categorification that exists in some  mathematical quarters. Such categorification may provide 
a formalism that may be able to bring forth unexpected aspects of non-extensive statistical mechanics, and touches upon on some aspects of topics 
discussed in this work. One application of this categorification that has touched upon Physics is that of Khovanov homology \cite{Khov, Witten} in 
relation to Chern-Simons theory and the Jones polynomial. 
It may also be worth studying the case of the Fukaya categories related to Lagrangian Floer cohomology in symplectic geometry \cite{Auroux} and 
mirror symmetry. In the context of entropy, alas only for $\mathcal{S}_{BGS}$ and in the spirit of categorification, one may appreciate some unique 
insights and viewpoints explored in the recent \cite{Gromov2, Gromov3} which may eventually turn out  to be particularly useful and illuminating. \\

%%%%%%%%%%%%%%%%%%%%%%%%%%%%%%%%%%%%%%%%%%%%%%%%%%%%%%%%%%%%%%%%%%%%%%%%%%%%% 

%%%%%%%%%%%%%%%%%%%%%%%%


\begin{thebibliography}{99} 


 \bibitem{Bal-P} Balian, R., \ Entropy: a Protean Concept, \ {\em S\'{e}m. Poincar\'{e}} \ {\bf 2003}, \ {\em 2}, \ 13-27.  
 \bibitem{Bal-Book} Balian, R. \ {\em From Microphysics to Macrophysics: Methods and Applications of Statistical Physics, \ Vol. 1}, \ 
                                Springer-Verlag, \ Berlin (1991).  
 \bibitem{Lesne} Lesne, A., \ Shannon Entropy: a rigorous notion at the crossroads between probability, information theory, dynamical 
                           systems and statistical physics, \ {\em Math. Struct. in Comp. Sci.}, \ {\bf 2014}, \ {\em 24 (3)}, \ 240311 (63 pages). 
 \bibitem{Coh} Cohen, E.G.D. \ Statistics and dynamics, \  {\em Physica A} \  {\bf 2002}, \ {\em 305} \  19-26.
 \bibitem{HC}  Havrda, J.;  Charv\'{a}t, F. \  Quantification method of classification processes. Concept of structural $\alpha$-entropy, \   
                                              {\em Kybernetika}  {\bf 1967}, \  {\em 3}, \ 30-35.
 \bibitem{Dar}  Z. Dar\'{o}czy,  \  Generalized Information Functions,  \ {\em Information and Control} \ {\bf 1970},  {\em 16}, \  36-51.
 \bibitem{CR} N.A. Cressie, T.R. Read, \  Multinomial goodness-of-fit tests, \  {\em J. Roy. Stat. Soc. B} \ {\bf 1984}, \ {\em 46}, \  440-464.     
 \bibitem{RC} Read, T.R., Cressie, N.A., \ {\em Goodness-of-fit statistics for discrete multivariate data}, \ Springer, \ New York, NY, USA  \ (1988).      
 \bibitem{T1} C. Tsallis, \  Possible generalisation of Boltzmann-Gibbs statistics, \ {\em J. Stat. Phys.} \ {\bf 1988} \ {\em 52}, \  479-487.
 \bibitem{T-book} C. Tsallis, \ \emph{Introduction to Nonextensive Statistical Mechanics: Approaching  
                            a Complex World}, \ Springer \ (2009).
  \bibitem{Kan1} Kaniadakis, G., \  Non-linear kinetics underlying generalized statistics, \ {\em Physica A} {\bf 2001}, \ {\em 296}, \ 405-425.
  \bibitem{Kan2} Kaniadakis, G. \   Statistical mechanics in the context of special relativity, \ {\em Phys. Rev. E} {\bf 2002}, \ {\em 66}, \ 056125.
  \bibitem{KanScar} Kaniadakis, G.; Scarfone, A.M., \ A new one-parameter defamation of the exponential function, \ {\em Physica A} \ {\bf 2002}, 
                                       \ {\em 305}, \ 69-75. 
  \bibitem{Kan3} Kaniadakis, G. \  Statistical mechanics in the context of special relativity II, \ {Em Phys. Rev. E} {\bf 2005}, \ {\em 72}, \ 036108. 
  \bibitem{Kan4} Kaniadakis, G. \ Theoretical Foundations and Mathematical Formalism of the Power-Law Tailed Statistical Distributions, \ 
                                                           {\em Entropy} {\bf 2013}, {\em 15}, 3983-4010.
 \bibitem{Nau1} Naudts, J. \ Deformed exponentials and logarithms in generalized thermostatistics, \ {\em Physica A} \ {\bf 2002}, \ {\em 316}, \ 323-334.  
 \bibitem{Nau2} Naudts, J.  \ Generalized Exponential Families and Associated Entropy Functions, \ {\em Entropy} \ {\bf 2008}, {\em 10}, 131-149.  
  \bibitem{Nau-book} Naudts, J., \ {\em Generalised Thermostatistics}, \ Springer-Verlag, \  London, \ UK \ (2011).
 \bibitem{NK1} Kalogeropoulos, N., \ Distributivity and deformation of the reals from Tsallis entropy, \ {\em Physica A} \ {\bf 2012}, \ {\em 391}, \ 1120-1127. 
 \bibitem{NK2} Kalogeropoulos, N., \ Tsallis entropy induced metrics and CAT(k) spaces, \ {\em Physica A} \ {\bf 2012}, \  {\em 391}, \ 3435-3445. 
 \bibitem{NK3} Kalogeropoulos, N., \ Vanishing largest Lyapunov exponent and Tsallis entropy, \ {\em QScience Connect} \ {\bf 2013}, \ {\em 2013}, \ 26.    
 \bibitem{NK4} Kalogeropoulos, N., \ Escort distributions and Tsallis entropy, \ {\sf arXiv:1206.5127}
 \bibitem{NK5} Kalogeropoulos, N., \ Tsallis entropy composition and the Heisenberg group, \ {\em Int. J. Geom. Methods Mod. Phys.} \ {\bf 2013}, 
                               \ {\em 10}, \ 1350032.
 \bibitem{NK6} Kalogeropoulos, N., \ Long-range interactions, doubling measures and Tsallis entropy, \ {\em Eur. Phys. Jour. B} \ {\bf 2014}, \ {\em 87}, 56. 
 \bibitem{NK7} Kalogeropoulos, N., \ Almost additive entropy, \ {\em Int. J. Geom. Methods Mod. Phys.} \ {\bf 2014}, \ {\em 11}, \ 1450040. 
 \bibitem{NK8} Kalogeropoulos, N., \ Groups, non-additive entropy and phase transitions, \ {\em Int. J. Mod. Phys. B} \ {\bf 2014}, \ {\em 28}, \ 1450162.
 \bibitem{NK9} Kalogeropoulos, N., \ Ricci curvature, isoperimetry and a non-additive entropy, \ {\em Entropy} \ {\bf 2015}, \ {\em 17}, \ 1278-1308. 
  \bibitem{Gor}  Gorban, A.N., \ Basic Types of Coarse-Graining, \  in {\em Model Reduction and Coarse-Graining Approaches for Multiscale Phenomena}, \ 
                         Gorban, A.N.; Kazantzis, N.; Kevrekidis, I.G.; \"{O}ttinger, H.C.; Theodoropoulos, C. \ (Eds.), \ Springer-Verlag, \ Berlin \ (2006);  \ pp. 117-176.                                    
 \bibitem{CFLV} Castiglione, P.; Falcioni, M.; Lesne, A.; Vulpiani, A., \ {\em Chaos and Coarse-Graining in Statistical Mechanics}, \ 
                                          Cambridge Univ. Press, \ Cambridge, \ UK \ (2008).  
 \bibitem{QQ} Quarati, F.; Quarati, P., \ Phase Space Cell in Nonextensive Classical Systems, \ {\em Entropy} \ {\bf 2003}, \ {\em 3}, \ 239-251.                      
 \bibitem{QL} Quarati, P.;  Lissia, M.,   \ The Phase Space Elementary Cell in Classical and Generalised Statistics, \ {\em Entropy} \ {\bf 2013}, 
                             \ {\em 15}, \ 4319-4333.
  \bibitem{FSV} Falasco, G.; Saggiorato, G.; Vulpiani, A., \ About the role of chaos and coarse-graining in statistical mechanics, \ {\em Physica A} \  {\bf 2015}, 
                             \ {\em 418}, \ 94-104.
  \bibitem{Coh2} Cohen, E.G.D., \ Boltzmann and Einstein: Statics and Dynamics - An unsolved problem, \ {\em Pramana} \ {\bf 2005}, \ {\em 64}, \ 635-643.                            
  \bibitem{Abe1} Abe, S.,  \ Essential discreteness in generalized thermostatistics with non-logarithmic entropy, \  {\em Europhys. Lett.}  {\bf 2010}, \ {\em 90} \ 
                                             50004.
  \bibitem{Andre}  Andresen, B.,  \ Comment on ``Essential discreteness  in generalised thermostats tics with non-logarithmic entropy" by Abe Sumiyoshi,  \   
                                                              {\em Europhys. Lett.} \ {\bf 2010}, \ {\em 92}, \ 40005. 
  \bibitem{Abe2} Abe, S., \  Reply to the Comment by B. Andresen, \ {\em  Europhys. Lett.} \ {\bf 2010}, \  {\em 92}, \ 40006.
  \bibitem{BOT} Bagci, G.B.;  Oikonomou, T.; Tirnakli, U., \ \ {\em Comment on ``Essential discreteness in
                                generalised thermostatistics with non-logarithmic entropy" by S. Abe,} \ \  {\sf arXiv:1006.1284}
   \bibitem{BL} Boon, J.P.; Lutsko, J.F., \  Nonextensive formalism and continuous Hamiltonian systems, \  {\em Phys. Lett. A} \ {\bf 2011}, \  {\em 375}, \ 
                                329-334. 
   \bibitem{LB} Lutsko, J.F.;  Boon, J.P., \ Questioning the validity of non-extensive thermodynamics for classical Hamiltonian systems, \  {\em Europhys. Lett.}  
                        \ {\bf 2011}, \ {\em 95}, \ 20006. 
   \bibitem{PR} Plastino, A.;  Rocca,, M.C.,  \ Possible divergences in Tsallis' thermostatstics, \ {\em Europhys. Lett.} \  {\bf 2013},  \ {\em 104}, \ 60003.
   \bibitem{WW} Wilk, G.;  W{\l}odarczyk, Z.,  \ Tsallis distribution with complex nonextensivity parameter q, \ {\em Physica A} \ {\bf 2014},  \  {\em 413}, \ 53-58.
   \bibitem{Touch} Touchette, H.,   \  When is a quantity additive, and when is it extensive?  \ {\em Physica A} \ {\bf 2002}, \  {\em 305}, \ 84-88.
   \bibitem{NLeMW} Nivanen, L.;  Le Mehaut\'{e}, A.; Wang, Q.A, \  Generalized algebra within nonextensive statistics, \ {\em Rep. Math. Phys.} \ 
                                                         {\bf 2003} \ {\em 52}, \ 437-444. 
   \bibitem{Borg}  Borges, E.P., \  A possible deformed algebra and calculus inspired in nonextensive thermostatistics,\ {\em Physica A} \ {\bf 2004}, 
                                                         \ {\em 340}, \  95-101.
   \bibitem{PLCPB} Petit Lob\~{a}o, T.C.;  Cardoso, P.G.S.;  Pinho, S.T.R.; Borges, E.P.,  \ Some properties of deformed q-numbers, \ 
                                                                           {\em Braz. J. Phys.} \ {\bf 2009}, \  {\em 39}, \ 402-407.
   \bibitem{Shan} C.E. Shannon, \  A mathematical theory of communication, \ {\em Bell. Syst. Tech. J.} \ {\bf 1948}, \  {\em 27}, \ 379-423. 
   \bibitem{Khin} Khinchin, A.Ya. \ The concept of entropy in the theory of probability, \ {\em Uspekhi Mat. Nauka} \ {\bf 1953}, \ {\em 8}, \ 3-20.
   \bibitem{Santos} Santos, R.J.V., \ Generalization of Shannon's theorem for Tsallis entropy, \ {\em J. Math. Phys.} \ {\bf 1997}, \ {\em 38}, \ 4104-4107.
   \bibitem{Abe3} Abe, S., \ Axioms and uniqueness theorem for Tsallis entropy, \ {\em Phys. Lett. A}  \ {\bf 2000}, \  {\em 271}, \ 74-79.
   \bibitem{Kac} Kac, V.; Cheung, P., \ {\em Quantum Calculus}, \ Springer-Verlag, New York, NY (2002).
   \bibitem{Livad} Livadiotis, G.; McComas, D.J.; \ Beyond kappa distributions: exploiting Tsallis statistical mechanics in space plasmas, \ 
                                                        {\em Jour. Geophys. Res.} \ {\bf 2009}, \ {\em 114}, \  A11105.
   \bibitem{PierLaz} Pierrard, V.; Lazar, M., \ Kappa distributions: theory and applications in space plasmas, \ {\em Solar Phys.} \ {\bf 2010}, \ 
                                    {\em 267}, \ 153-174.  
   \bibitem{BomHenSor} Bombelli, L.; Henson, J.; Sorkin, R.D., \ Discreteness without symmetry breaking: A theorem, \ {\em Mod. Phys. Lett. A} \ 
                                      {\bf 2009}, \ {\em 24}, \ 2579-2587.
   \bibitem{NK10}  Kalogeropoulos, N., \ Weak Chaos from Tsallis Entropy, \  {\em QScience Connect} \ {\bf 2012}, \ {\em 2012}, \ 12.                                 
   \bibitem{CFKP} Cannon, J.W.; Floyd, W.J.; Kenyon, R.; Parry, W.R., \ Hyperbolic Geometry, \ in \ {\em Flavors of Geometry}, \ Levy, S.
                                                       (Ed.), \ MSRI Publications, {\em Vol. 31}, \ Cambridge Univ. Press, \ Cambridge, UK \ (1997); \ pp. 59-115.
   \bibitem{Tal} Talagrand, M.; \ A new look at independence, \ {\em Ann. Probab.} \ {\bf 1996}, \ {\em 24}, \ 1-34.
   \bibitem{Triebel} Triebel, H., \ {\em Theory of Function Spaces}, \ Monographs in Mathematics, Vol. 78, \ Birkh\"{a}user-Verlag, \ Basel, \ (1983).  
   \bibitem{GorKOT} Gorban, A.N.; Karlin, I.V.; \"{O}ttinger, H.C.;  Tatarinova, L.L., \ Ehrenfests' arguments extended to a formalism of nonequilibrium 
                                             thermodynamics, \ {\em Phys. Rev. E} \ {\bf 2001}, \ {\em 63}, \ 066124.  
    \bibitem{GorKarl} Gorban, A.N.; Karlin, I.V., \ Uniqueness of thermodynamic projector and kinetic basis of molecular individualism, \ {\em Physica A} 
                                           \ {\bf 2004}, \ {\em 336}, \ 391-432. 
    \bibitem{GellMH1} Gell-Mann, M.;  Hartle, J.B.,  \ Quasiclassical coarse-graining and thermodynamic entropy, \  {\em Phys. Rev. A} \ 
                                                       {\bf 2007}, \ {\em 76}, \ 022104.
    \bibitem{GellMH2} Gell-Mann, M.; Hartle, J.B.,  \ Adaptive coarse graining, environment, strong decoherence, and quasi classical realms,  
                                                        \    {\em Phys. Rev. A} \ {\bf 2014}, \ {\em 89}, \ 052125.                                    
    \bibitem{vanKamp} van Kampen, N.G. \ {\em Stochastic Processes in Physics and Chemistry}, \ North Holland, \ Amsterdam, \ The Netherlands (1992).
    \bibitem{Smale1} Smale, S., \ Structurally stable systems are not dense, \ {\em Amer. J. Math.} \ {\bf 1966}, \ {\em 88}, \ 491-496.
    \bibitem{Smale2} Smale, S. \ Differentiable Dynamical Systems, \ {\em Amer. Math. Soc. Bull.} \  {\bf 1967}, \ {\em 73}, \ 747-817.
     \bibitem{KatokHas}  Katok, A.; Hasselblatt, B., \ \emph{Introduction to the Modern Theory of Dynamical Systems}, \  Cambridge Univ. Press, 
                                                \ Cambridge, UK, \ (1995).  
    \bibitem{ANPS} Arendt, W.; Nittka, R.; Peter, W.; Steiner, F., \ Weyl's Law: Spectral Properties of the Laplacian in Mathematics and Physics, \ in \
                                        {\em Mathematical Analysis of Evolution, Information and Complexity}, \ Arendt, W.; Schleich, W.P. (Eds.), \ Wiley-VCH Verlag, \ Weinheim, 
                                                     Germany, \ (2009); \ pp. 1-71.                                           
    \bibitem{Nash1} Nash, J., \  $C^1$ isometric imbeddings, \ {\em Ann. Math.} \ {\bf 1954}, \ {\em 60}, \ 383-396.   
    \bibitem{Nash2} Nash, J., \  The imbedding Problem for Riemannian Manifolds, \ {\em Ann.Math.} \ {\bf 1956}, \ {\em 63}, \ 20-63.  
    \bibitem{Sakai}  Sakai, T.,  \ {\em Riemannian Geometry}, \ Transl. Math. Monogr., \ Vol. 149, \  Amer. Math. Soc., \ Providence, \ RI \ (1996).
    \bibitem{Gromov1} Gromov, M., \ Sign and Geometric Meaning of Curvature, \ {\em Rendinconti del Seminario Matematico e Fisico di Milano}, \
                                                                {\bf 1991}, \ {\em 61}, \ 9-123.   
    \bibitem{Gromov-book} Gromov, M., \ {\em Metric Structures for Riemannian and Non-Riemannian Spaces}, \  Birkh\"{a}user, \ Boston, \ MA  \ (1999).
    \bibitem{Arn} Arnold, V.I., \  {\em Mathematical Methods of Classical Mechanics}, 2nd Ed., \ Graduate Texts in Math. Vol. 60, Springer-Verlag, NY (1978). 
    \bibitem{HoferZehn} H. Hofer, E. Zehnder, \ {\em Symplectic Invariants and Hamiltonian Dynamics}, \ Birkh\"{a}user-Verlag, \ Basel, \ Switzerland \  (1994). 
    \bibitem{McDSal} McDuff, D.; Salamon, D., \ {\em Introduction to Symplectic Topology}, 2nd Ed., \ Clarendon Press, \ Oxford, UK, \ (1998).
    \bibitem{Polt} Polterovich, \ {\em The Geometry of the Group of Symplectic diffeomorphisms}, \ Lect. in Math., \ ETH Z\"{u}rich, \ Birkh\"{a}user Verlag, \
                             Basel, \  Switzerland, \ (2001).
    \bibitem{Schlenk} Schlenk, F., \ {\em Embedding problems in symplectic geometry}, \ de Gruyter Expositions In Math. Vol. 40, \ Berlin, \ Germany \ (2005).                         
    \bibitem{Zehn} Zehnder, E., {\em Lectures on Dynamical systems; Hamiltonian Vector Fields and Symplectic Capacities}, \ Eur. Math. Soc., \ Z\"{u}rich, \ 
                                                                                   Switzerland \ (2010).     
    \bibitem{Vit-Book} Viterbo, C., \ {\em An Introduction to Symplectic Topology through Sheaf Theory}, \ available online at \  
                                                                    {\sf \  \ http://www.math.polytechnique.fr/cmat/viterbo/Eilenberg/Eilenberg.pdf}
    \bibitem{Eliash} Eliashberg, Y., \ Symplectic Topology in the nineties, \ {\em Diff. Geom. Appl.} \ {\bf 1998}, \ {\em 9}, \ 59-88. 
    \bibitem{CHLS} Cielebak, K.; Hofer, H.; Latschev, J.; Schlenk, F., \ Quantitiative Symplectic Geometry, \ in \ {\em Dynamics, Ergodic Theory and Geometry},
                                    \ B. Hasselblatt \ (Ed.), \ MSRI Publ. Vol. 54, \ Cambridge Univ. Press, \ Cambridge, UK (2007); \ pp. 1-44. 
    \bibitem{McDuff} McDuff, D. \ Symplectic Topology Today, \ Colloquium Lectures, \ Joint Mathematical Meetings, \  Baltimore,  \ January 2014, \  
                              available online at \  {\sf  http://jointmathematicsmeetings.org/meetings/national/jmm2014/colloqnov2.pdf} 
    \bibitem{Gromov-PDR} Gromov, M., \ {\em Partial Differential Relations}, \ Springer-Verlag, \ Berlin \  (1986).
    \bibitem{Gromov-Pseudo} Gromov, M., \ Pseudoholomorphic curves in symplectic manifolds, \ {\em Invent. Math.} \ {\bf 1985}, \ {\em 82}, \ 307-347. 
    \bibitem{Gallav} Gallavotti, G., \ {\em Statistical Mechanics: A Short Treatise}, \ Springer-Verlag, \ Berlin \ (1999).
    \bibitem{Katok} Katok, A. B., \ Ergodic perturbations of degenerate integrable Hamiltonian systems, \ {\em Math. USSR, Izvestija} \ {\bf 1973}, \ {\em 7}, 535-571. 
    \bibitem{deGos-Luef} de Gosson, M.; Luef, F., \ Symplectic capacities and the geometry of uncertainty: The irruption of symplectic topology in 
                                                                     classical and quantum  mechanics, \ {\em Phys. Rep.} \ {\bf 2009}, \ {\em 484}, 131-179. 
    \bibitem{deGos1} de Gosson, M., \ The symplectic camel and phase space quantization, \ {\em J. Phys. A: Math. Gen.} \ {\bf 2001}, \ {\em 34}, \ 10085-10096.
    \bibitem{deGos2} de Gosson, M., \ The ``symplectic camel principle" and semiclassical mechanics, \ {\em J. Phys. A: Math. Gen.} \ {\bf 2002}, \ {\em 35}, \ 
                                                                               6825-6851. 
    \bibitem{deGos3} de Gosson, M., \ Phase Space quantisation and the Uncertainty Principle, \ {\em Phys. Lett. A} \ {\bf 2003}, \ {\em 317}, \ 365-369.
    \bibitem{deGos4} de Gosson, M., \ Symplectically Covariant Schr\"{o}dinger Equation in Phase Space, \ {\em J. Phys. A: Math. Gen} \ {\bf 2005}, \  {\em 38}, \ 
                                                                              9263-9287.
    \bibitem{deGos5} de Gosson, M., \ The Symplectic Camel and the Uncertainty Principle: The Tip of an Iceberg? \ {\em Found. Phys.} \ {\bf 2009}, \ {\em 39}, \ 
                                                                               194-214.
    \bibitem{deGos6} de Gosson, M.A.; Hiley, B.J.,  \ Imprints of the Quantum World in Classical Mechanics, \ {\em Found. Phys.} \ {\bf 2011}, \ {\em 41}, \ 1415-1436.
    \bibitem{deGos7} de Gosson, M., \ Quantum Blobs, \ {\em Found. Phys.}, \ {\bf 2013}, \ {\em 43}, \ 440-457. 
    \bibitem{deGos8} de Gosson, M., \ The symplectic egg in quantum and classical mechanics, \ {\em Amer. J. Phys.} \ {\bf 2013}, \ {\em 81}, \ 328-337.
    \bibitem{deGos-Book} de Gosson, M.A., \ {\em Symplectic Methods in Harmonic Analysis and Mathematical Physics}, \ Birkh\"{a}user, \ 
                                                                                         Basel, \ Switzerland \ (2011). 
    \bibitem{Taubes-Book} Taubes, C.H. \ {\em Seiberg-Witten and Gromov invariants for symplectic 4-manifolds}, \ International Press, \ Boston, \  (2005). 
    \bibitem{EkelHofer1} Ekeland, I.; Hofer, H., \ Symplectic Topology and Hamiltonian Dynamics I, \ {\em Math. Zeit.} \ {\bf 1990}, \ {\em 200}, \ 355-378. 
    \bibitem{EkelHofer2} Ekeland, I., Hofer, H., \ Symplectic Topology and Hamiltonian Dynamics II, \ {\em Math. Zeit.} \ {\bf 1990}, \ {\em 203}, \ 553-567. 
    \bibitem{Hofer} Hofer, H. \  On the topological properties of symplectic maps, \ {\em Proc. Roy. Soc. Edin., Sect. A} \ {\bf 1990}, \ {\em 115}, \ 25-38.  
    \bibitem{HZ-Cap} Hofer, H.; Zehnder, E.; \ A new capacity for symplectic manifolds, \ in \ {\em Analysis, et cetera: Research Papers Published in Honor 
                                        of J\"{u}rgen Moser's 60th Birthday}, \ Rabinowitz, P.H.; Zehnder, E. (Eds.), \  Academic Press, \ San Diego, \ USA \ (1990); \ pp. 405-428.  
    \bibitem{Hutch} Hutchings, M., \ Quantitative Embedded Contact Homology, \ {\em J. Diff. Geom.} \ {\bf 2011}, \ {\em 88}, \ 231-266. 
    \bibitem{Fef} Fefferman, C., \ The uncertainty principle, \ {\em Bull. Amer. Math. Soc. (N.S.)} \ {\bf 1983}, \ {\em 9}, \ 129-206. 
    \bibitem{Fol-Sit} Folland, G.B.; Sitaram, A., \ The uncertainty principle: A mathematical survey, \  {\em Jour. Four. Anal. Appl.} \ {\bf 1997}, \ {\em 3}, \ 207-238.
    \bibitem{Sib} Siburg, K.F., \ Symplectic capacites in two dimensions, \ {\em Manuscr. Math.} \ {\bf 1993}, \ {\em 78}, \ 149-163.
    \bibitem{Jiang} Jiang, M.-Y., \ Hofer-Zehnder symplectic capacity for two-dimensional manifolds, \ {\em Proc. Roy. Soc.  Edin.} \ {\bf 1993}, \ {\em 123A}, \ 945-950. 
    \bibitem{Fed} Federer, H., \ {\em Geometric Measure Theory}, \ Springer-Verlag, \ Berlin, \  Germany (1969).
    \bibitem{McDuff-Sal} McDuff, D.; Salamon, D., \ {\em $J$-Holomorphic Curves and Symplectic Topology}, \  Amer. Math. Soc. Colloq. Publ., \ 
                                                                   Providence, \ USA \ (2004). 
    \bibitem{Vit} Viterbo, C., \ Viterbo, C., \ Metric and isoperimetric problems in symplectic geometry, \ {\em J. Amer. Math. Soc.} \ {\bf 2000}, \ {\em 13}, \ 411-431.
    \bibitem{Art-Ostr} Artstein-Avidan, S.; Ostrover, Y., \ On Symplectic Capacities and Volume Radius, \ available online at \ 
                                                                         {\sf http://arxiv.org/pdf/math.SG/0603411.pdf} 
    \bibitem{Art-Mil-Ostr} Artstein-Avidan, S.; Milman, V.; Ostrover, Y., \ The M-ellipsoid, symplectic capacities and volume, \ {\em Comment. Math. Helv.} \ {\bf 2008}, \ 
                                                                          {\em 83}, \ 359-369.     
     \bibitem{Hermann} Hermann, D., \ Non-equivalence of symplectic capacities for open sets with restricted contact type boundary, \ Orsay Preprint No. 32, \ 
                              (29/4/1998). 
   \bibitem{Rock} Rockafellar, R.T., \ {\em Convex Analysis}, \ Princeton Univ. Press, \ Princeton, NJ \ (1970).
   \bibitem{Lind-Jafr} J. Lindenstrauss, L. Tzafriri, \ {\em Classical Banach Spaces I and II}, \ Springer-Verlag, \ Berlin,  (1977).
  \bibitem{Mil-Sch} Milman, V.D.; Schechtman, G., \ {\em Asymptotic Theory of Finite Dimensional Normed Spaces: With an Appendix by M. Gromov}, 
                                                                            \ Lect. Notes Math., \  Vol. 1200, \ Springer-Verlag, \ Berlin \ (1986). 
  \bibitem{Pis} Pisier, G., \ {\em The Volume of Convex Bodies and Banach Space Geometry}, \ Camb. Tracts in Math. Vol. 94, \ Camb. Univ. Press, \ 
                                                                                              Cambridge,  \ UK  \ (1989).                                                                                                                                                                        
  \bibitem{Benya} Benyamini, Y.; Lindenstrauss, J., \ {\em Geometric Nonlinear Functional Analysis: Volume 1}, \ Coll. Publ., Vol 48, \ Amer. Math. Soc., \ 
                                                                                                         Providence,\  RI \ (2000).   
  \bibitem{Schn} Schneider, R., \ {\em Convex Bodies: The Brunn-Minkowski Theory}, \ 2nd Expanded Ed., \  Encycl. Math. Appl. Vol. 151, \ 
                                                                                          Camb. Univ. Press, \ Cambridge, \ UK \ (2014).                                                                                                      
  \bibitem{AAGM} Artstein-Avidan, S.; Giannopoulos, A.; Milman, V., \ {\em Asymptotic Geometric Analysis, Part I}, \ Math. Surv.  Mon., Vol. 202, \ 
                                                                                                         Amer. Math. Soc., \ Providence, \ RI \ (2015). 
  \bibitem{Ball} Ball, K., \ An Elementary Introduction to Modern Convex Geometry, \ in \ {\em Flavors of Geometry}, Levy, S. (Ed.). \ 
                                          Math. Sci. Res. Inst. Publ., Vol. 31, \ Cambridge Univ. Press,  Cambridge, UK  (1997).
  \bibitem{Versh} Vershynin, R., \ Lectures in Geometric Functional Analysis, \ available online at \   
                                                 {\sf http://www-personal.umich.edu/~romanv/papers/GFA-book/GFA-book.pdf    }
  \bibitem{GM1} Giannopoulos, A.A.; Milman, V.D., \ Euclidean Structure in Finite Dimensional Normed Spaces, \ in \ {\em Handbook of the 
                                              Geometry of Banach Spaces, Vol. 1}, \ Johnson, W.B.; Lindenstrauss, J. (Eds.), \ Elsevier Science, \
                                                                    Amsterdam, \  The Netherlands \  (2001); \ pp. 707-779.  
  \bibitem{GM2} Giannopoulos, A.A.; Milman, V.D., \ Asymptotic Convex Geometry: a short overview, \ in \ {\em Different Faces of Geometry}, \ 
                                               Donaldson, S.K.; Eliashberg, Y.; Gromov, M., (Eds.), Kluwer Academic / Plenum Publishers, \ New York, NY (2004); \  pp. 87-162.
   \bibitem{Ruelle}  Eckmann, J.-P.; Ruelle, D., \ Ergodic theory of chaos and strange attractors, \ {\em Rev. Mod. Phys.} \ {\bf 1985}, \ {\em 57}, \ 617-656.                                            
   \bibitem{Young} Young, L.-S., \  What are SRB Measures, and Which Dynamical Systems Have Them?, \ {\em J. Stat. Phys.} \ {\bf 2002}, \ {\em 108}, \ 
                                                                                                                           733-754.   
  \bibitem{Mil1} Milman, V.D., \ A new proof of the theorem of A. Dvoretzky on sections of convex bodies, \ {\em Funct. Anal. Appl.} \ {\bf 1971}, \ {\em 5}, \ 28-37.
  \bibitem{Mil2} Milman, V.D., \ Asymptotic properties of functions of several variables defined on homogeneous spaces, \ {\em Sov. Math. Dokl.}  \ {\bf 1971}, 
                                                                           \ {\em 12}, \ 1277-1281.
  \bibitem{GroMil1} Gromov, M.; Milman, V.D., \ A Topological Application of the Isoperimetric Inequality, \ {\em Amer. Jour. Math.} \ {\bf 1983}, \ {\em 105}, \ 843-854.
  \bibitem{GroMil2} Gromov, M.; Milman, V.D.; \ Generalization of the spherical isoperimetric inequality to uniformly convex Banach spaces, \ {\em Comp. Math.} \ 
                                                                                  {\bf 1987}, \ {\em 62}, \ 263-282.
  \bibitem{Mil3} Milman, V.D., \ The heritage of P. L\'{e}vy in geometrical functional analysis, \ {\em Ast\'{e}risque} \ {\bf 1988}, \ {\em 157-158}, \ 273-301.                                                                              
  \bibitem{Ledoux} Ledoux, M. \ {\em The Concentration of Measure Phenomenon}, \ Math. Surveys Monogr. \ Vol. 89, \ Amer. Math. Soc., \ Providence, RI  \ (2001). 
  \bibitem{GW} Gaspard, P.; Wang, X.J., \ Noise, chaos and the ($\epsilon, \tau$) entropy per unit time, \ {\em Phys. Rep.} \ {\bf 1993}, \ {\em 235}, \ 291-343.    
  \bibitem{BBH} Budd, C.J.; Butler, R.; Hunt, G.W., \ Geometry and mechanics of layered structures and materials, \ {\em Phil. Trans. R. Soc. A} \ {\bf 2012}, \ 
                                                                                        {\em 370 (1965)}, \  1721- 2026.
  \bibitem{Hor} Ho\v{r}ava, P., \ Quantum Gravity at a Lifshitz point, \ {\em Phys. Rev. D} \ {2009}, \ {\em 79}, \ 084008.                                                                                                                                                                  
  \bibitem{Chavel} Chavel, I., \ {\em Isoperimetric Inequalities: Differential and Analytic Perspectives}, \ Cambridge Tracts Math. Vol. 145, \ Cambridge Univ. Press, \
                                                Cambridge, \ UK \ (2001).
  \bibitem{Blaschke} Blaschke, W., \  \"{U}ber Affine Geometrie VII: Neue Extremeigenschaften von Ellipse und Ellipsoid, \ {\em Ber. Verh. S\"{a}chs. Acad. Wiss. 
                                                                 Leipzig, Math.-Phys. KL}  \ {\bf 1917}, \ {\em 69}, \ 412-420.      
  \bibitem{Santalo} Santal\'{o}, L.A., \ Un invariate afin para los cuerpos convexos de espacio de n dimensiones, \ {\em Portugaliae Math.} \ {\bf 1949}, \ 
                                                                        {\em 8}, \ 155-161.
  \bibitem{Petty} Petty, C.M., \  Affine isoperimetric problems, \ in \ {\em Discrete geometry and convexity}, \ Ann. New York Acad. Sci., Vol. 440, \
                                                                 New York Acad. Sci., \  New York, NY \  (1985); \ pp. 113-127. 
 \bibitem{Kuper} Kuperberg, G., \ From the Mahler conjecture to Gauss linking integrals, \ {\em Geom. Funct. Anal.} \ {\bf 2008}, \ {\em 18}, \ 870-892.
 \bibitem{Art-Kar-Ost} Artstein-Avidan, S.; Karasev, R.; Ostrover, Y., \ From Symplectic Measurements to the Mahler Conjecture, \ {\em Duke Math. Jour.} \ 
                                                                         {\bf 2014}, \ {\em 163}, \ 2003-2022.
  \bibitem{Ostrover} Ostrover, Y., \ When symplectic topology meets Banach space geometry, \ in \ {\em Proceedings of the International Congress of 
                                           Mathematicians, Seoul 2014, Vol II}, \ Jang, S.Y.; Kim, Y.R., Lee, D.-W.; Yie, I. (Eds.), \ Kyung Moon SA Co. Ltd, \ Seoul, \ Korea \ 
                                            (2014); \  pp.  959-981.     
  \bibitem{Polch} Polchinski, J., \ String Theory, \ Volume I and Volume II \ Cambridge Univ. Press, \ Cambridge, \ UK \ (1998).
  \bibitem{HKKP} Hori, K.; Katz, S.; Klemm, A.; Pandharipandhe, R.; Thomas, R.; Vafa, C.; Vakil, R.; Zaslow, E., \ {\em Mirror Symmetry}, \  
                                       Clay. Math. Inst., \ Vol. 1, \  Amer. Math. Soc., \  Providence, RI  \ (2003).
  \bibitem{Ram} Ramallo, A.V., \ {\em Introduction to the AdS/CFT Correspondence}, \ available as \ {\sf arXiv:1310.4319v3}   
  \bibitem{Polyakov} Polyakov, A.M., \  {\em Gauge Fields and Strings}, Harwood Academic Publ., \ Chur, \  Switzerland \ (1987). 
  \bibitem{UBB1} Urmossy, K.; Barnaf\"{o}ldi, G.G.; Bir\'{o}, T.S., \ Generalized Tsallis statistics in electron-positron collisions, \ {\em Phys. Lett. B} \ {\bf 2011}, \ 
                                                                                         {\em 701}, \ 111-116. 
  \bibitem{UBB2} Urmossy, K.; Barnaf\"{o}ldi, G.G.; Bir\'{o}, T.S., \ Microcanonical jet-fragmentation in proton-proton collisions at LHC energy, \ {\em Phys. Lett. B} \ 
                                                                           {\bf 2012}, \ {\em 718}, \ 125-129.
  \bibitem{Dep} Deppman, A., \ Self-consistency in non-extensive thermodynamics of highly excited harmonic states, \ {\em Physica A} \ 
                                                  {\bf 2012}, \ {\em 391}, \ 6380-6385.   
  \bibitem{WW1} Wilk, G.; W{\l}odarczyk, Z.; Self-similarity in jet events following from p-p collisions at LHC, \ {\em Phys. Lett. B} \ {\bf 2013}, \ {\em 727}, \ 163-167.
  \bibitem{MCD} Marques, L.; Cleymans,J.; Deppman, A., \ Description of High-energy pp Collisions Using Tsallis Thermodynamics: Transverse Momentum and 
                                                               Rapidity Distributions, \ available at \ {\sf arXiv:1501.00953}  
  \bibitem{WWCT} Wong, C.-Y.; Wilk, G; Cirto, L.J.L.; Tsallis, C., \ From QCD-based hard scattering to nonextensive statistical mechanical descriptions of transverse 
                                            momentum spectra in high energy $pp$ and $p\bar{p}$ collisions, \ {\em Phys. Rev. D.} \ {\bf 2015}, \ {\em 91}, \ 114027. 
  \bibitem{Khov} Khovanov, M., \ A Categorification of the Jones Polynomial, \ {\em Duke Math. Jour.} \ {\bf 2000}, {\em 101}, 359-426.
  \bibitem{Witten} Witten, E., \ Khovanov Homology and Gauge Theory, \ in \ {\em Proceedings of the FreedmanFest}, \ Kirby, R.; Krushkal, V.; Wang, Z. (Eds.), 
                                                       Geom. Topol. Mon., Vol. 18, \ Math. Sci. Publ.\  (2012); \  pp. 291-308.  
  \bibitem{Auroux} Auroux, D., \ A beginner's introduction to Fukaya categories, \ in \ {\em Contact and Symplectic Topology}, \ Bourgeois F.; Colin, V.; Stipsicz, A. 
                                  (Eds.),  \ Bolyai Soc. Math. Studies Vol. 26, \  Springer, \ Heidelberg, \ Germany \ (2014).
   \bibitem{Gromov2} Gromov, M., \ {\em In Search for a Structure, Part 1: On Entropy} \ available at \ 
                                                             {\sf http://www.ihes.fr/~gromov/PDF/structre-serch-entropy-july5-2012.pdf}   
   \bibitem{Gromov3} Gromov, M., \ {\em Six Lectures on Probability, Symmetry, Linearity, October 2014, Jussieu (Unedited)}, \ available at \ \ 
                                            {\sf http://www.ihes.fr/~gromov/PDF/probability-huge-Lecture-Nov-2014.pdf}   
\end{thebibliography}
\end{document}